\documentclass[12pt,eadjointtfnpsf]{article}

\usepackage{amsmath,amsbsy,amssymb,latexsym,epsfig}

\def\parity{\mathfrak R}

\newcommand{\BE}{\begin{equation}}
\newcommand{\EE}{\end{equation}}
\newcommand{\BEA}{\begin{eqnarray}}
\newcommand{\EEA}{\end{eqnarray}}

\def\12{\frac{1}{2}}
\def\bea{\begin{eqnarray}}
\def\eea{\end{eqnarray}}
\def\ba{\begin{array}}
\def\ea{\end{array}}

\def\one-loop{\mbox{\scriptsize one-loop}}

\def\G{\Gamma}

\catcode`\@=11

\@addtoreset{equation}{section}
\def\theequation{\arabic{section}.\arabic{equation}}

\def\@normalsize{\@setsize\normalsize{15pt}\xiipt\@xiipt
\abovedisplayskip 14pt plus3pt minus3pt%
\belowdisplayskip \abovedisplayskip
\abovedisplayshortskip \z@ plus3pt%
\belowdisplayshortskip 7pt plus3.5pt minus0pt}

\def\small{\@setsize\small{13.6pt}\xipt\@xipt
\abovedisplayskip 13pt plus3pt minus3pt%
\belowdisplayskip \abovedisplayskip
\abovedisplayshortskip \z@ plus3pt%
\belowdisplayshortskip 7pt plus3.5pt minus0pt
\def\@listi{\parsep 4.5pt plus 2pt minus 1pt
\itemsep \parsep
\topsep 9pt plus 3pt minus 3pt}}

\def\underline#1{\relax\ifmmode\@@underline#1\else
$\@@underline{\hbox{#1}}$\relax\fi}
\@twosidetrue

\relax

\catcode`@=12

\catcode`\@=11

\def\section{\@startsection{section}{1}{\z@}{3.5ex plus 1ex minus
.2ex}{2.3ex plus .2ex}{\large\bf}}

\def\thesection{\Roman{section}.}

\def\appendix{\setcounter{section}{0}
\def\thesection{Appendix }

\def\theequation{\Alph{section}.\arabic{equation}}}

\catcode`\@=12

\relax

\def\figcap{\section*{Figure Captions\markboth
{FIGURECAPTIONS}{FIGURECAPTIONS}}\list
{Fig. \arabic{enumi}:\hfill}{\settowidth\labelwidth{Fig. 999:}
\leftmargin\labelwidth
\advance\leftmargin\labelsep\usecounter{enumi}}}
 \relax
\def\tablecap{\section*{Table Captions\markboth
{TABLECAPTIONS}{TABLECAPTIONS}}\list
{Table \arabic{enumi}:\hfill}{\settowidth\labelwidth{Table 999:}
\leftmargin\labelwidth
\advance\leftmargin\labelsep\usecounter{enumi}}}
 \relax
\def\reflist{\section*{References\markboth
{REFLIST}{REFLIST}}\list
{[\arabic{enumi}]\hfill}{\settowidth\labelwidth{[999]}
\leftmargin\labelwidth
\advance\leftmargin\labelsep\usecounter{enumi}}}
 \relax

\newskip\humongous \humongous=0pt plus 1000pt minus 1000pt

\newif\ifdtup

\def\tr{\mathop{\rm tr}}

\def\Im{\mathop{\rm Im}}
\def\Re{\mathop{\rm Re}}

\def\beq{\begin{equation}}
\def\eeq{\end{equation}}

\def\beqn{\begin{eqnarray}}
\def\eeqn{\end{eqnarray}}
\relax

\def\G2{{\; \rm GeV/}c2}
\def\G{\; \rm GeV}

\def\dotx{\dotx{\dot\overline{x}}}

\relax

\newcommand\CL{{\mathcal L}}
\newcommand\CW{{\mathcal W}}
\newcommand\CF{{\mathcal F}}
\newcommand\CD{{\mathcal D}}
\newcommand\fD{{\mathfrak D}}

\newcommand\CN{{\mathcal N}}

\newcommand\CV{{\mathcal V}}

\newcommand\CM{{\mathcal M}}

\def\llangle{\langle \hspace{-1mm} \langle}
\def\rrangle{\rangle \hspace{-1mm} \rangle}

\renewcommand{\thefootnote}{\fnsymbol{footnote}}
\def\fnote#1#2{\begingroup\def\thefootnote{#1}\footnote{#2}\addtocounter
{footnote}{-1}\endgroup}
\begin{document}
%
%
\begin{titlepage}

\begin{flushright}
\normalsize
March, 2005 \\
OCU-PHYS 228 \\
hep-th/0503113 \\
\end{flushright}

\begin{center}
{\large\bf  
Partial Breaking of $\mathcal{N}=2$ Supersymmetry \\
 and of Gauge Symmetry \\
 in the $U(N)$ Gauge Model }
\end{center}

\vfill

\begin{center}
{%
K. Fujiwara$^a$\footnote{e-mail: fujiwara@sci.osaka-cu.ac.jp}
\quad, \quad
H. Itoyama$^a$\footnote{e-mail: itoyama@sci.osaka-cu.ac.jp}
\quad and \quad
M. Sakaguchi$^b$\footnote{e-mail: msakaguc@sci.osaka-cu.ac.jp}
}
\end{center}

\vfill

\begin{center}
$^a$ \it Department of Mathematics and Physics,
Graduate School of Science\\
Osaka City University\\
\medskip

$^b$ \it Osaka City University Advanced Mathematical Institute
(OCAMI)

\bigskip

3-3-138, Sugimoto, Sumiyoshi-ku, Osaka, 558-8585, Japan \\

\end{center}

\vfill

\begin{abstract}

We explore vacua of the $U(N)$ gauge model with $\mathcal{N}=2$ supersymmetry
recently constructed in hep-th/0409060.
In addition to the vacuum previously found with unbroken $U(N)$ gauge symmetry 
in which $\mathcal{N}=2$ supersymmetry is partially broken to $\mathcal{N}=1$, 
we find cases in which the gauge symmetry is broken to a product gauge group
$\displaystyle{\prod _{i=1}^n U(N_i)}$. 
The $\mathcal{N}=1$ vacua are selected by the requirement of a positive 
definite K\"ahler metric.
We obtain the masses of the supermultiplets appearing on the 
$\mathcal{N}=1$ vacua.

\end{abstract}

\vfill

\setcounter{footnote}{0}
\renewcommand{\thefootnote}{\arabic{footnote}}

\end{titlepage}
\section{Introduction}

This is the sequel to our previous papers
\cite{FIS1,FIS2} and intends to investigate further properties
of our model arising from an interplay between 
$\mathcal{N}=2$ supersymmetry and nonabelian gauge symmetry.
In ref \cite{FIS1}, we have successfully constructed 
the $\mathcal{N}=2$ supersymmetric $U(N)$ gauge model in four spacetime
dimensions, generalizing the abelian self-interacting model
given some time ago in ref \cite{APT}.
The gauging of $U(N)$ isometry associated with the special K\"ahler
geometry, and the discrete $\parity$  symmetry are  the
primary ingredients of our construction.
The model spontaneously breaks $\mathcal{N}=2$ supersymmetry to 
$\mathcal{N}=1$. The second supersymmetry realized on the broken
phase acts as an approximate fermionic $U(1)$ shift symmetry.
This, combined with the notion of prepotential as an input function,
tells that the model should be interpreted as a low energy effective
action (LEEA) designed to apply microscopic calculation
invoking spectral Riemann surfaces\cite{SW,SWINT},
matrix models (see \cite{Morozov:2005mz} for a recent review) 
and/or string theory\cite{string}
to various physical processes.  
Although we will not investigate in this paper,
it is interesting to try to find the origin of and the role played by
the three parameters of our model $e, m,$ and $\xi$ in the developments
beginning with the work of \cite{DV,DVint,Cachazo:2002ry}. 
Connection to various
compactification schemes of strings, branes and M theory\cite{Witten:1995ex}
is anticipated and some work along this direction has already
appeared \cite{KP}. 

Partial spontaneous breaking of extended supersymmetries
appears not possible  from the consideration of the algebra among
the  supercharges.
The basic mechanism enabling the partial breaking is in fact a 
modification of the local version of the extended supersymmetry
algebra by an additional spacetime independent term which
forms a matrix with respect to extended indices
\cite{Lopuszanski:1978df,central charge}:
\beqn
\left\{ \bar{Q}_{\dot{\alpha}}^j,\mathcal{S}_{\alpha i}^m (x) \right\}
=2(\sigma^n)_{\alpha \dot{\alpha}} \delta_{i}^{\ j}
T_n^m(x)+(\sigma^m)_{\alpha \dot{\alpha}} C_i^{\ j}~.
 \label{currentalgebra}
\eeqn
Note that this last term is not a vacuum expectation value  but
simply follows from the algebra of the extended supercurrents and from that
the triplet of
the auxiliary fields $\boldsymbol{D}^a$ is complex undergoing 
the algebraic constraints of \cite{N=2 constraint}.
 
Our model predicts
\beqn
  C_i^{\ j}=+4m\xi (\boldsymbol{\tau}_1)_i^{\ j}. \label{Cij}
\eeqn
Separately, we find that the scalar potential 
takes a vacuum expectation value
\beqn
  \llangle\CV\rrangle=\mp 2m\xi =2| m\xi |. 
\eeqn
(See (\ref{potential4.14}).)
Once these are established, it is easy to see that partial breaking of
extended supersymmetries is a reality: after ninety degree rotation,
the vacuum annihilates half of  the supercharges  while the remaining
half takes nonvanishing and in fact infinite 
($\sim |m\xi| \int d^4 x$) matrix elements.

The thrust of the present paper is to explore vacua of the model 
in which the $U(N)$ gauge symmetry is spontaneously broken to 
various product gauge groups and to compute the mass spectrum 
on the $\mathcal{N}=1$ vacua  as a function of input data, 
which are the prepotential derivatives and $e, m$ and $\xi$.
In the next section, we review the $\mathcal{N} =2$ supersymmetric  
$U(N)$ gauge model. 
Analysis of the vacua is given in section III
and we determine the vacuum expectation value of the auxiliary
fields $\boldsymbol{D}_{a}$. 
We also collect some properties of the derivatives
of the prepotential. In section IV, we exhibit the
Nambu-Goldstone fermion of the model and the $\mathcal{N}=1$ vacua
are selected by the requirement of the positivity of the K\"ahler
metric. In section V,  we show that the vacua permit various breaking
patterns of the $U(N)$ gauge group into product gauge groups
$\displaystyle{\prod_{i=1}^N U(N_i)}$. 
The ``triplet-doublet splitting'' at $N=5$
is discussed.
Finally in section VI, 
we compute the masses of the bosons
and of the fermions of the model and obtain three types of
$\mathcal{N}=1$ (on-shell) supermultiplets.
In the Appendix, we collect some formulas associated with the
standard basis of the $u(N)$ Lie algebra.

It will be appropriate to introduce here notation to label 
the generators of 
$u(N)$ Lie algebra by indices.
A set of $u(N)$ generators is first labelled by indices 
$a,b,...\ =0,1,...\ N^2-1$.
Here $0$ refers to the overall $U(1)$ generator.
In a basis in which the decoupling of $U(1)$ is manifest, 
we label the generators
belonging to the maximal Cartan subalgebra by $i,j,k,...$ 
while the generators belonging to the 
roots (the non-Cartan generators) are labelled by $r,s,...\ $\ .
In our analysis of vacua, 
we employ the standard basis of the $u(N)$ Lie algebra.
See the Appendix for this basis.
The diagonal generators in this basis are labelled by 
$\underline{i},\underline{j},\underline{k},...$ and are 
referred to as those in the eigenvalue basis.
The non-Cartan generators associated with unbroken gauge symmetry
are labelled  by $r',s',...$ while the remaining non-Cartan 
generators representing broken gauge symmetry are labelled by 
$\mu, \nu,...$\ .
Our physics output, 
mass spectrum of our model only distinguishes 
the broken generators from the unbroken ones.
We therefore introduce $\alpha,\beta,\gamma,...$\ 
as a union of $\underline{i},\underline{j},\underline{k},...$ 
and $r,s,...$ in order to 
label the entire unbroken generators.
Our final formula will be expressible in terms of 
$\alpha, \beta,...$ and $\mu,\nu,...$ only.

\section{Review of the $U(N)$ gauge model}

The $\CN=2$ $U(N)$ gauge model constructed in \cite{FIS1}
is composed of a set of $\CN=1$ chiral multiplets $\Phi=\Phi^at_a$
and a set of $\CN=1$ vector multiplets $V=V^at_a$,
where $N\times N$ hermitian
 matrices $t_a$, $(a=0,\ldots N^2-1)$,
  generate $u(N)$, $[t_a, t_b]=if^c_{ab}t_c$.
These superfields, $\Phi^a$ and $V^a$, contain 
component fields $(A^a,\psi^a,F^a)$ and $(v_m^a,\lambda^a,D^a)$,
respectively.
This model is described by an analytic function 
$\CF(\Phi)$.\fnote{$\flat $}{$\CF_a\equiv \partial_a\CF$ and 
$\CF_{ab}\equiv \partial_a\partial_b\CF$}
The kinetic term of  $\Phi$
is given by the K\"ahler potential
$K(\Phi^a,\Phi^{*a})=\frac{i}{2}(\Phi^a\CF_a^*-\Phi^{*a}\CF_a)$
of the special K\"ahler geometry
as $\CL_{K}=\int d^2\theta^2 d\bar\theta^2 K(\Phi^a,\Phi^{*a})$.
The K\"ahler metric 
$g_{ab^*}\equiv\partial_a\partial_{b^*}K(A^a,A^{*a})=\Im \CF_{ab}$
admits  isometry $U(N)$.
The $U(N)$ gauging 
is accomplished 
\cite{Bagger:1982fn,Hull:1985pq}
by
adding 
to $\CL_K$
$\CL_\Gamma$
which is specified by
the Killing potential
$\fD_a=
-ig_{ab}f^b_{cd}A^{*c}A^d$.
The kinetic term of $V$
is given as 
$\CL_{\CW^2}=-\frac{i}{4}\int d^2\theta^2\CF_{ab}\CW^a\CW^b+c.c$,
where $\CW^a$ is the gauge field strength of $V^a$.
This model contains the superpotential term
$\CL_W=\int d\theta^2 W +c.c$ with $W=eA^0+m\CF_0$,
and the Fayet-Iliopoulos D-term 
$\CL_D=\sqrt{2}\xi D^0$ as well \cite{FI} .
Gathering these together, the total Lagrangian
of the $\CN=2$ $U(N)$ model
is given as
\begin{eqnarray}
\CL&=&\CL_K+\CL_\Gamma+\CL_{\CW^2}+\CL_W+\CL_D~.
\label{L}
\end{eqnarray}
Eliminating the auxiliary fields 
 by using their equations of motion
\begin{eqnarray}
D^{a}&=&
\hat{D}^a
-\frac{1}{2}g^{ab}\left(
\fD_b + 2\sqrt{2}\xi\delta_{b}^0
\right)
~,~~~
\hat{D}^a\equiv
-\frac{\sqrt{2}}{4}g^{ab} 
\left( 
\CF_{bcd}
\psi^d\lambda^c
+
\CF_{bcd}^*
\bar\psi^d\bar\lambda^c \right)
~, 
\label{D}
\\
F^a&=&\hat{F}^a
-g^{ab^*}\partial_{b^*} W^*
~,~~~
\hat{F}^a\equiv
\frac{i}{4}g^{ab^*} 
\left( 
\CF_{bcd}^*
\bar\lambda^c\bar\lambda^d
-
\CF_{bcd}
\psi^c\psi^d \right)
~,
\label{F}
\end{eqnarray}
the Lagrangian $\CL$ (\ref{L}) takes the following form:
\begin{eqnarray}
\label{L'}
\CL'
&=&
\CL_{\rm{kin}}
+\CL_{\rm{pot}}
+\CL_{\rm{Pauli}}
+\CL_{\rm{mass}}
+\CL_{\rm{fermi^4}}
~,
\end{eqnarray}
with
\begin{eqnarray}
\CL_{\rm{kin}}&=&
-g_{ab^*}\mathcal{D}_m A^a \mathcal{D}^m A^{*b}
-\frac{1}{4} g_{ab} v_{mn}^a v^{bmn}
-\frac{1}{8} \Re (\CF_{ab})\, \epsilon^{mnpq}v_{mn}^a v_{pq}^b
\nonumber\\&&
+ \left[
-\frac{1}{2} \CF_{ab} \lambda^a \sigma^m \mathcal{D}_m
\bar{\lambda}^b
-
\frac{1}{2}\CF_{ab}
\psi^a\sigma^m \mathcal{D}_m \bar{\psi}^b
~~+~~c.c.\right]~,
\label{L: kin} \\
\CL_{\rm{pot}}&=&
-
g^{ab}
 \left(
\frac{1}{8}\fD_{a}
\fD_b
+ \xi^2\delta_{a}^0\delta_{b}^0
\right)
-g^{ab^*}\partial_a W \partial_{b^*} {W}^*,
\label{L: pot}
\\
\CL_{\rm{mass}}&=&
\left[
-\frac{i}{4}g^{cd^*}
\CF_{abc}
\partial_{d^*} W^*
 (\psi^a\psi^b+\lambda^a\lambda^b)
~~+c.c. \right]
\nonumber\\&&
+
\left[
\frac{1}{2\sqrt{2}}\left(
 g_{ac^*}k_b^*{}^{c}-g_{bc^*}k_a^*{}^{c}
 -\sqrt{2}\xi\delta_{c}^0
g^{cd}
\CF_{abd}
 \right) \psi^a\lambda^b
~~+c.c. \right] ~,
\label{L:mass}
\end{eqnarray}
where we have defined the covariant derivative by
$\CD_m\Psi^a\equiv\partial_m\Psi^a-\frac{1}{2}f^a_{bc}v_m^b\Psi^c$
for $\Psi^a\in\{A^a, \psi^a,\lambda^a\}$,
and 
$v^a_{mn}\equiv\partial_mv_n^a-\partial_nv_m^a-\frac{1}{2}f^a_{bc}v_m^bv_n^c$.
The holomorphic Killing vectors
$k_a=k_a{}^b\partial_b$
are
generated by the Killing potential
$\fD_a$
as
$k_a{}^b=-ig^{bc^*}\partial_{c^*}\fD_a$
and
$k^*_a{}^b=ig^{b^*c}\partial_{c}\fD_a$.
Here, we have omitted
$\CL_{\textrm{Pauli}}$ and $\CL_{\rm{fermi^4}}$
as they are irrelevant for our purposes in this paper.

In \cite{FIS1}, Lagrangians, $\CL$ and $\CL'$, are
shown to be invariant under the $\parity$-action,
\begin{eqnarray}
\parity: 
\begin{array}{c}
\left(
  \begin{array}{c}
    \lambda^a   \\
    \psi^a   \\
  \end{array}
\right)
\to
\left(
  \begin{array}{c}
    \psi^a  \\
    -\lambda^a 
  \end{array} 
\right) 
~,~~~
  \xi \to -\xi
  \end{array}
  ~,
\label{extended R}
\end{eqnarray} 
and for $\CL'$ in addition 
\begin{eqnarray}
\parity:
\begin{array}{c}
F^a+g^{ac^*}\partial_{c^*} W^*
\to
F^{*b}+g^{db^*}\partial_{d}W~, \\
D^c+\frac{1}{2}g^{cd}\fD_d
\to
-( D^c+\frac{1}{2}g^{cd}\fD_d
)~.
\end{array}
\end{eqnarray}
This property guarantees the $\CN=2$ supersymmetry of the model
as follows.
By construction, the action, $S\equiv \int \CL$ or $\int \CL'$,
is invariant under the $\CN=1$
supersymmetry, $\delta_1 S=0$.
Operating the $\parity$-action on this equation, one finds
$0=\parity \delta_1S\parity^{-1}=\parity \delta_1\parity^{-1}S$.
This implies that $S$ is invariant under the second supersymmetry
$\delta_2\equiv\parity \delta_1\parity^{-1}$
as well
in addition to the first supersymmetry $\delta_1$.
We have also given in \cite{FIS1} (see Appendix A)
a proof of $\mathcal{N}=2$ supersymmetry of our action,
using the canonical transformation acting only on the fields
without invoking $\xi \rightarrow -\xi$.
The $\CN=2$ supersymmetry transformations
are obtained by covariantizing the $\mathcal{N}=1$ transformations with 
respect to $\parity$.
Using the doublet of fermions
\begin{eqnarray}
\boldsymbol{\lambda}_I^{\ a} &\equiv& 
\left(
\begin{array}{c}
\lambda^a \\
\psi^a
\end{array}
\right)
~,~~~
\boldsymbol{\lambda}^{Ia} \equiv
 \epsilon ^{IJ} \boldsymbol{\lambda}_J^{\ a} 
~,~~~
\bar{\boldsymbol{\lambda}}^{Ia} \equiv
\left(
\begin{array}{c}
\bar{\lambda}^a \\
\bar{\psi}^a
\end{array}
\right)
~,
~~~
\bar{\boldsymbol{\lambda}}_I^{\ a} \equiv
 \epsilon_{IJ} \bar{\boldsymbol{\lambda}}^{Ja}
~,
\end{eqnarray}
and
the doublet of supersymmetry transformation parameters
\begin{eqnarray}
\boldsymbol{\eta}_I\equiv
\left(
  \begin{array}{c}
  \eta_1     \\
  \eta_2     \\
  \end{array}
\right)
~,~~~
\boldsymbol{\eta}^I
\equiv \epsilon^{IJ}\boldsymbol{\eta}_J
~,~~~
\bar{\boldsymbol{\eta}}^J \equiv
\left(
\begin{array}{c}
\bar{\eta}_1 \\
\bar{\eta}_2
\end{array}
\right)
~,~~~
\bar{\boldsymbol{\eta}}_J \equiv 
\epsilon_{JI} \bar{\boldsymbol{\eta}}^I \ ,
\end{eqnarray}
where $\epsilon^{IJ}$ 
is given by
$\epsilon^{12}=\epsilon_{21}=1$ and
$\epsilon^{21}=\epsilon_{12}=-1$,
the $\mathcal{N}=2$ transformations are written as
\begin{eqnarray}
\boldsymbol{\delta} A^a &=&
\sqrt{2} \boldsymbol{\eta}_J \boldsymbol{\lambda}^{Ja}, \label{four} \\
\boldsymbol{\delta} v_m^{a} &=&
i \boldsymbol{\eta}_J \sigma_m \bar{\boldsymbol{\lambda}}^{Ja}   
-i \boldsymbol{\lambda}_J^{\ a}{\sigma}_m\bar{\boldsymbol{\eta}}^J \ ,
\label{delta A} \\
\boldsymbol{\delta \lambda}_J^{\ a} &=& 
(\sigma^{mn} \boldsymbol{\eta}_J)v_{mn}^{a}
+\sqrt{2}i(\sigma^m \bar{\boldsymbol{\eta}}_J) \mathcal{D}_m A^a
+i(\boldsymbol{\tau} \cdot \boldsymbol{D}^a)_J{}^{K} \boldsymbol{\eta}_K
-\frac{1}{2} \boldsymbol{\eta}_J f^a_{\ bc} A^{*b} A^c
~. 
\ \label{delta lambda} 
\end{eqnarray}
Here, $\boldsymbol{D}^a$ represent the three-vectors
\begin{eqnarray}
\boldsymbol{D}^a &=&
\hat {\boldsymbol{D}}^a -\sqrt{2} g^{ab^*} 
\partial_{b^*}
\left( \boldsymbol{\mathcal{E}}A^{*0}+\boldsymbol{\mathcal{M}}
{\mathcal{F}}_0^* \right),
\label{3-vector D} \\
\hat{\boldsymbol{D}}^a&=&
(\sqrt{2}\Im \hat F^a,-\sqrt{2}\Re \hat F^a, \hat D^a), 
\\
\boldsymbol{\mathcal{E}}&=&(0,\ -e,\ \xi)~,~~~
\boldsymbol{\mathcal{M}}=(0,\ -m,\ 0), 
\label{E & M}
\end{eqnarray}
and $\boldsymbol{\tau}$ are the Pauli matrices. 
Let us also note that 
\begin{eqnarray}
\Im \boldsymbol{D}^a = \delta^a_{\ 0} (-\sqrt{2} m)
\left(
\begin{array}{c}
0 \\
1 \\
0
\end{array}
\right).
\label{ImD}
\end{eqnarray}
This simply follows from (\ref{tildeD}) 
and also a consequence from the superspace constraints 
derived in \cite{N=2 constraint}.

The construction of the extended supercurrents is somewhat involved 
as is fully discussed in \cite{FIS1} 
(See \cite{Ferrara:1974pz,Itoyama:1996hh} for this).
Nonetheless it is easy to extract the piece contributing to 
$C_i^{\ j}$ in the algebra (\ref{currentalgebra}).
It comes from the structure of
\begin{eqnarray}
\boldsymbol{\tau} \cdot \boldsymbol{D}^{*b} \boldsymbol{\tau} 
\cdot \boldsymbol{D}^a = 
\boldsymbol{D}^{*b} \cdot \boldsymbol{D}^a \boldsymbol{1} 
+ i 
\left( \boldsymbol{D}^{*b} \times \boldsymbol{D}^a \right) 
\cdot \boldsymbol{\tau}.
\end{eqnarray}
The second term is nonvanishing only for complex 
$\boldsymbol{D}^a$ and this is the piece responsible for $C_i^{\ j}$.
After using (\ref{3-vector D}) and (\ref{ImD}), we derive (\ref{Cij}).


\section{Analysis of vacua}

Let us examine the scalar potential of the model,
$\CV=-\CL_{\rm pot}$,
\begin{eqnarray}
\CV
&=& g^{ab}\left(
\frac{1}{8}\fD_a\fD_b
+\xi^2\delta_{a}^0\delta_{b}^0
+\partial_a W\partial_{b^*} W^*
\right) \nonumber \\
&=& g^{ab} \left( \frac{1}{8} \fD_a \fD_b +
\partial_a \left( \boldsymbol{\mathcal{E}} A^0 
 + \boldsymbol{\mathcal{M}} 
\mathcal{F}_0 \right)
\cdot
\partial_{b^*} 
\left( \boldsymbol{\mathcal{E}} A^0 
 + \boldsymbol{\mathcal{M}} \mathcal{F}_0 \right)^* \right).
\label{scalarpotential}
\end{eqnarray}
For our present purpose, 
it is more useful to convert this expression into
\begin{eqnarray}
\mathcal{V}=
\frac{1}{8}g_{bc} \mathfrak{D}^b \mathfrak{D}^c 
+\frac{1}{2}g^{bc} \tilde{\boldsymbol{D}}_b^* 
\cdot \tilde{\boldsymbol{D}}_c~,
\label{scalarpotential2}
\end{eqnarray}
where
\begin{eqnarray}
\mathfrak{D}^a&=&g^{ab}\fD_b=
-i f^a_{cd}A^{*c} A^d~,
\end{eqnarray}
and
\begin{eqnarray}
\tilde{\boldsymbol{D}}_b &\equiv&
g_{ba} \boldsymbol{\tilde D}^a
=
-\sqrt{2}
\partial_{b^*}
\left( \boldsymbol{\mathcal{E}} A^{*0} 
 + \boldsymbol{\mathcal{M}} 
\mathcal{F}_0^* \right)
= \sqrt{2}
\left(
\begin{array}{c}
0 \\
\partial_{b^*} W^* \\
-\xi \delta_b^{\ 0}
\end{array}
\right) 
~,
\label{tildeD}
\end{eqnarray}
as well as (\ref{ImD}).
We derive from (\ref{scalarpotential2})-(\ref{tildeD}),
\begin{eqnarray}
\frac{\partial \mathcal{V}}{\partial A^a}&=&
\frac{1}{4} g_{bc} 
\partial_a \mathfrak{D}^b
\mathfrak{D}^c
+\frac{1}{8} 
\partial_a g_{bc}
\mathfrak{D}^b \mathfrak{D}^c
+\frac{i}{4} \mathcal{F}_{abc} \tilde{\boldsymbol{D}}^{*b} 
\cdot \tilde{\boldsymbol{D}}^{c}
-\frac{\sqrt{2}}{2} 
\mathcal{F}_{abc} \boldsymbol{\mathcal{M}}\delta^b_0 
\cdot \tilde{\boldsymbol{D}}^{c}
\nonumber \\
&=& \frac{1}{4} g_{bc} 
\partial_a \mathfrak{D}^b
\mathfrak{D}^c
+\frac{1}{8} 
\partial_a g_{bc}
\mathfrak{D}^b \mathfrak{D}^c+\frac{i}{4} \mathcal{F}_{abc} 
\tilde{\boldsymbol{D}}^b \cdot \tilde{\boldsymbol{D}}^c.
\label{vacuum}
\end{eqnarray}
This is an expression valid in any vacuum and generalizes the one 
found in \cite{FIS1} (eq.(4.17))
for the unbroken vacuum.

In order to examine general vacua,
let indices $a=(i,r)$ and $i\, (r)$ 
label the (non) Cartan generators of $u(N)$ (see the Appendix).
We are interested in the vacuum at which $\langle A^{r}\rangle=0$.
The vacuum condition (\ref{vacuum})
reduces \cite{FIS2} to
\begin{eqnarray}
0=
\frac{i}{4}\langle \CF_{abc}\boldsymbol{D}^b\cdot \boldsymbol{D}^c\rangle , 
\label{vacuum condition}
\end{eqnarray}
because $\langle\fD^a\rangle=\langle -if^a_{ij}A^{*i}A^j\rangle=0$.
Here and henceforth, we drop the tilde on $\boldsymbol{D}^a$
noting that
$\langle \boldsymbol{\tilde D}^a\rangle =
\langle \boldsymbol{D}^a\rangle$ as the vacuum expectation value.

It is more convenient
to work on the set of bases 
(the eigenvalue bases)
in which the Cartan subalgebra of $u(N)$ is spanned by 
\begin{equation}
(t_{\underline{i}}
)_j^{\ k} = \delta_{\underline{i}}^{\ k} \delta_j^{\ \underline{i}} \ .
\ \ \ \ \ \ \ \ \ 
(\underline{i}={1} \sim {N}~ ,\ \ \ \ 
j,k=1 \sim N~.)
\end{equation}
These $t_{\underline{i}}$ correspond to $H_{\underline{i}}$
in the Appendix.
Let us introduce a
matrix
which transforms
the standard bases labelled by $i,j,k$ of the Cartan generators into 
the eigenvalue bases labelled by $\underline{i},\underline{j},\underline{k}$
as
\begin{eqnarray}
t_{\underline{i}}&=&O_{\underline{i}}{}^{{j}}t_{{j}}~,~~~
\label{O:1}
\\
t_i&=&O_i{}^{\underline{j}}t_{\underline{j}}~.~~~
\label{O:2}
\end{eqnarray}
This matrix satisfies
$O_i{}^{\underline{j}} O_{\underline{j}}{}^k=\delta_i{}^k$ and $
O_{\underline{i}}{}^j O_j{}^{\underline{k}} =
\delta_{\underline{i}}{}^{\underline{k}}$.
We normalize the standard $u(N)$ Cartan generators $t_i$ as
$\tr ( t_it_j )=\frac{1}{2}\delta_{ij}$,
which implies that the overall $u(1)$ generator is
$t_0=\frac{1}{\sqrt{2N}}{\bold 1}_{N\times N}$.
Let us derive useful relations which will be exploited in what follows.
Summing up equation (\ref{O:1}) with respect to $\underline{i}$,
we find
$\displaystyle{\sqrt{2N}t_0=\sum_{\underline{j}}O_{\underline{j}}{}^it_i}$,
and thus
$\displaystyle{\sum_{\underline{j}}O_{\underline{j}}{}^i=\sqrt{2N}\delta_0{}^i}$.
Taking the trace of (\ref{O:1}),
we obtain 
\begin{eqnarray}
O_{\underline{i}}{}^0=\sqrt{\frac{2}{N}} \ .
\end{eqnarray}
Taking the trace of (\ref{O:2}), we obtain 
$\displaystyle{\sqrt{\frac{N}{2}}\delta_i^{\ 0}=
\sum_{\underline{j}} O_i^{\ \underline{j}}}$.
It follows from ($\ref{O:2}$) with $i=0$ that 
\begin{eqnarray}
O_0^{\ \underline{j}}=\frac{1}{\sqrt{2N}} \ . \label{Ooj}
\end{eqnarray}
On the other hand, non-Cartan generators $t_r$ are
$E_{\underline{i}\underline{j}}^\pm=
\pm E_{\underline{j}\underline{i}}^\pm$
in the Appendix.
Thus $\Phi$ can be expanded as 
$
\displaystyle{
\Phi=
\sum_i \Phi^{{i}}t_{i}
+\sum_r \Phi^r t_r=
\sum_{\underline{i}} \Phi^{\underline{i}}t_{\underline{i}}
+\frac{1}{2}
\sum_{
\underline{i},\underline{j}
(\underline{i} \neq \underline{j})
}
\left(
\Phi^{\underline{i}\underline{j}}_+E_{\underline{i}\underline{j}}^+
+\Phi^{\underline{i}\underline{j}}_-E_{\underline{i}\underline{j}}^-
\right)
}
$
with 
$\Phi^{\underline{i}\underline{j}}_\pm=
\pm \Phi^{\underline{j}\underline{i}}_\pm$.
In the Appendix, we explain our notation in some detail.

Let us collect some properties of $\langle\CF_{ab}\rangle$
and $\langle\CF_{abc}\rangle$ for the following $\CF$:
\begin{eqnarray}
\CF=\sum_{\ell=0}^k\frac{C_\ell}{\ell!}\tr\Phi^\ell.
\label{prepot}
\end{eqnarray}
We note that the matrix $\langle \Phi \rangle$ 
is complex and normal and the eigenvalues $\lambda^{\underline{i}}$ 
are in general complex.
Letting $\langle\Phi\rangle=\lambda^{\underline{i}}t_{\underline{i}}$,
we find that the nonvanishing $\langle \mathcal{F}_{ab} \rangle$ are
\begin{eqnarray}
\langle\CF_{\underline{i}\underline{i}}\rangle&=&
\sum_\ell\frac{C_\ell}{(\ell-2)!}
( \lambda^{\underline{i}})^{\ell-2}~,\\
\langle\CF_{\pm \underline{i}\underline{j},\pm \underline{i}\underline{j}}
\rangle&\equiv&\langle
\frac{\partial^2\CF}
{\partial \Phi^{\underline{i}\underline{j}}_\pm
\partial \Phi^{\underline{i}\underline{j}}_\pm}
\rangle
=
\left\{
  \begin{array}{ll}\displaystyle
\sum_{\ell}\frac{C_\ell}{2(\ell-1)!}
\frac{(\lambda^{\underline{i}})^{\ell-1}-
(\lambda^{\underline{j}})^{\ell-1}}{\lambda^{\underline{i}}-
\lambda^{\underline{j}}}       
&\mbox{if}~~ \lambda^{\underline{i}}\neq \lambda^{\underline{j}} \ ,    \\
\displaystyle
\sum_{\ell}\frac{C_\ell}{2(\ell-2)!}
(\lambda^{\underline{i}})^{\ell-2}       
&\mbox{if}~~ \lambda^{\underline{i}}= \lambda^{\underline{j}} \ .    \\
  \end{array}
\right.
\end{eqnarray}
See the Appendix .
These imply that $\langle g_{ab}\rangle$ is diagonal:
$\langle g_{\underline{i}\underline{j}}\rangle\propto 
\delta_{\underline{i}\underline{j}}$,
$\langle g_{rs}\rangle\propto 
\delta_{rs}$ and
$\langle g_{ir}\rangle=\langle g^{ir}\rangle=0$.
In addition, 
we have 
$\langle\CF_{+ \underline{i}\underline{j},+ \underline{i}\underline{j}}\rangle
=\langle\CF_{- \underline{i}\underline{j},- \underline{i}\underline{j}}\rangle$,
i.e. $\langle g_{+ \underline{i}\underline{j},+ \underline{i}\underline{j}}\rangle
=\langle g_{- \underline{i}\underline{j},- \underline{i}\underline{j}}\rangle$.
In particular we note that,
for directions $\Phi^{\underline{i}\underline{j}}_\pm$
with $\lambda^{\underline{i}}=\lambda^{\underline{j}}$,
$\langle\CF_{\pm \underline{i}\underline{j},\pm \underline{i}\underline{j}}\rangle
= \frac{1}{2} \langle\CF_{\underline{i}\underline{i}}\rangle$,
i.e.
$\langle g_{\pm \underline{i}\underline{j},\pm \underline{i}\underline{j}}\rangle
=\frac{1}{2}\langle g_{\underline{i}\underline{i}}\rangle$.
For $\langle\CF_{abc}\rangle$, the nonvanishing components are
\begin{eqnarray}
\langle\CF_{\underline{i}\underline{i}\underline{i}}\rangle&=&
\sum_\ell\frac{C_\ell}{(\ell-3)!}
( \lambda^{\underline{i}})^{\ell-3}~,\\
\langle\CF_{\underline{k},\pm \underline{i}\underline{j},
\pm \underline{i}\underline{j}}\rangle&=&
\left\{
  \begin{array}{ll}\displaystyle
\sum_{\ell}\frac{C_\ell}{2(\ell-1)!}
(\delta_{\underline{i}\underline{k}}
\frac{\partial}{\partial\lambda^{\underline{i}}}
+\delta_{\underline{j}\underline{k}}
\frac{\partial}{\partial\lambda^{\underline{j}}})
\frac{(\lambda^{\underline{i}})^{\ell-1}-(\lambda^{\underline{j}})^{\ell-1}}
{\lambda^{\underline{i}}-\lambda^{\underline{j}}}       
&\mbox{if}~~ \lambda^{\underline{i}}\neq \lambda^{\underline{j}} \ ,   \\
\displaystyle
\sum_{\ell}\frac{C_\ell}{2(\ell-3)!}
\delta_{\underline{i}\underline{k}}
(\lambda^{\underline{i}})^{\ell-3}       
&\mbox{if}~~ \lambda^{\underline{i}}= \lambda^{\underline{j}} \ .  \\
  \end{array}
\right.~~~
\end{eqnarray}
Obviously
$\partial_{\underline{i}}
\langle\CF_{abc\cdots  }\rangle=
\langle\partial_{\underline{i}}
\CF_{abc\cdots  }\rangle$.
Finally, we note here remarkable relations
\begin{eqnarray}
\langle\CF_{0,\pm \underline{i}\underline{j},
\pm \underline{i}\underline{j}}\rangle&=&
\left\{
  \begin{array}{ll}
\displaystyle
\sum_{\ell}\frac{C_\ell}{2\sqrt{2N}(\ell-2)!}
\frac{(\lambda^{\underline{i}})^{\ell-2}-(\lambda^{\underline{j}})^{\ell-2}}
{\lambda^{\underline{i}}-\lambda^{\underline{j}}}
=\frac{\langle\CF_{\underline{i}\underline{i}}\rangle
-\langle\CF_{\underline{j}\underline{j}}\rangle
}{2\sqrt{2N}(\lambda^{\underline{i}}-\lambda^{\underline{j}})}        
&\mbox{if}~~ \lambda^{\underline{i}}\neq \lambda^{\underline{j}}~,     \\
\displaystyle
\frac{1}{2\sqrt{2N}}\langle\CF_{\underline{i}\underline{i}\underline{i}}\rangle
&\mbox{if}~~ \lambda^{\underline{i}}= \lambda^{\underline{j}}~,    \\
  \end{array}
\right.
\nonumber\\
\label{key relation}
\end{eqnarray}
which play a key role in the analysis of the mass spectrum.

Let us return to the vacuum condition  (\ref{vacuum condition}).
This condition
is automatically satisfied for $a=r$ because
 $\langle\boldsymbol{D}^r\rangle=-\sqrt{2}\langle g^{rs}
 (\boldsymbol{\mathcal E}\delta_s^0+\boldsymbol{\mathcal M}\CF_{0s}^*)\rangle=0$.
Noting that
the only nonvanishing second and third derivatives
for the
Cartan directions 
are
the diagonal ones, namely, 
$\mathcal{F}_{\underline{i} \underline{i}}$ and $\mathcal{F}_{\underline{i} \underline{i} \underline{i}}$,
the vacuum condition  (\ref{vacuum condition})
reduces to
\begin{equation}
\left< \mathcal{F}_{\underline{j} \underline{j} \underline{j}} 
\boldsymbol{D}^{\underline{j}} \cdot \boldsymbol{D} ^{\underline{j}} \right>=0
\ \ \ \ \ \ \ \ 
\textrm{with $\underline{j}$ not summed},
\ \ 
1\leq \underline{j} \leq N \ . \label{VacuumConditon}
\end{equation}
The points specified by 
$\langle\CF_{\underline{j}\underline{j}\underline{j}}\rangle=0$
are not stable vacua because
$\langle\partial_{\underline{j}}\partial_{\underline{j}^*}\CV\rangle=0$.
At the stable vacua, we obtain 
$\left< \boldsymbol{D}^{\underline{j}} \cdot
\boldsymbol{D}^{\underline{j}} \right>=0$, 
or equivalently
\begin{equation}
\left< \boldsymbol{D}_{\underline{j}} \cdot
\boldsymbol{D}_{\underline{j}} \right>=0 \ ,
\end{equation}
where
\begin{equation}
\langle\boldsymbol{D}_{\underline{i}}\rangle=
 O_{\underline{i}}{}^{j} \langle\boldsymbol{D}_{{j}}\rangle=
\sqrt{2}
\left(
\begin{array}{c}
0 \\
\langle\frac{\partial}{\partial \lambda^{* \underline{i}}} W^*\rangle 
\\
\langle
- \xi \frac{\partial}{\partial \lambda^{*\underline{i}}} 
(A^{*0})
\rangle
\end{array}
\right)
=
\sqrt{2}
\left(
\begin{array}{c}
0 \\
eO_{\underline{i}}{}^0+mO_0{}^{\underline{i}}
\langle\CF^*_{\underline{i}\underline{i}}\rangle
\\
-\xi O_{\underline{i}}{}^0
\end{array}
\right)
~.
\end{equation}
We have determined
the vacuum expectation values 
of the following quantities:
\begin{eqnarray}
m \llangle\mathcal{F}^*_{\underline{j} \underline{j}} \rrangle
&=& - \frac{O_{\underline{j}}{}^0}{O_0{}^{\underline{j}} } (
e  \mp i \xi) =-2(e  \mp i \xi), 
\label{Fii:vacuum} \\
\llangle g_{\underline{j} \underline{j}}\rrangle &=&
 \mp 2\frac{\xi}{m}, \\
\llangle \boldsymbol{D}_{\underline{j}} \rrangle &=&
2 \frac{\xi}{\sqrt{N}}
\left(
\begin{array}{c}
0 \\
\pm i \\
-1
\end{array}
\right)\equiv \boldsymbol{d}_{(\pm)}, \\
\llangle \boldsymbol{D}^{\underline{j}}\rrangle 
&=& \frac{m}{\sqrt{N}}
\left(
\begin{array}{c}
0 \\
-i \\
\pm 1
\end{array}
\right)
\equiv \boldsymbol{d}^{(\pm)}. 
\label{3-vector: d}
\end{eqnarray}
We use $\llangle \cdots \rrangle$ for those vacuum expectation values 
which  are determined as the solutions
to (\ref{VacuumConditon}).
Note that the sign factor $\pm$ can be chosen freely for each 
$\underline{j}$.
Let $\boldsymbol{\mathcal{M}}_+$ be the set chosen from 
$1 , \ldots , N$
in which the $+$ sign is chosen in (\ref{3-vector: d})
and 
let $\boldsymbol{\mathcal{M}}_-$ be the one in which
the $-$ sign is chosen. 
We will see shortly that this determines the number of 
spontaneously broken supersymmetries.
As one sees from (\ref{L: kin}),
$\llangle \Im \CF_{ab}\rrangle$ measures the inverse of the 
squared coupling constant $\frac{1}{g_{YM}^2}$,
while $\llangle \Re \CF_{ab}\rrangle$
is the $\theta$-angle
of QCD.

\section{Partially broken supersymmetry and NG fermions.} \label{PBSsection}

Let us examine the supersymmetry transformation of fermions (\ref{delta lambda}),
which reduces at the vacuum to 
\begin{eqnarray}
\llangle \boldsymbol{\delta\lambda}^a_I\rrangle
=i\llangle ({\boldsymbol \tau}\cdot {\boldsymbol D}^a)_I{}^J\rrangle
\boldsymbol{\eta}_J~.
\label{susy}
\end{eqnarray}
As
$\llangle \boldsymbol{D}^r\rrangle=0$,
the fermion $\boldsymbol{\lambda}_I^r$ on the vacuum 
is invariant under the supersymmetry
\begin{eqnarray}
\llangle\boldsymbol{\delta}\boldsymbol{\lambda}_I^r\rrangle=0
~.
\end{eqnarray}
The Nambu-Goldstone fermion signaling supersymmetry breaking
is contained in $\boldsymbol{\lambda}_I^{{\underline{i}}}$.
For $\boldsymbol{\delta}\boldsymbol{\lambda}_I^{{\underline{i}}}$,
the $2\times 2$ matrix $\boldsymbol{\tau}\cdot \boldsymbol{D}^{{\underline{i}}}$
is easily diagonalized as
\begin{eqnarray}
\llangle
\delta\left(
\frac{\lambda^{\underline{i}}\pm\psi^{\underline{i}}}{\sqrt{2}}
\right)\rrangle
=\mp \frac{1}{\sqrt{2}}\llangle D_2^{\underline{i}}\mp{{i}}D_3^{\underline{i}}\rrangle
(\eta_1\mp \eta_2)~.
\end{eqnarray}

If either 
$\boldsymbol{\mathcal{M}}_-$ or $\boldsymbol{\mathcal{M}}_+$
is empty, 
$\llangle \boldsymbol{D}^{\underline{j}} \rrangle$ 
$(\underline{j}=1,...,  N)$
we have obtained are all identical.
When $\boldsymbol{\mathcal{M}}_-$ is empty,
the matrix 
$\boldsymbol{\tau}\cdot \boldsymbol{D}^{\underline{j}}
=\boldsymbol{\tau}\cdot \boldsymbol{d}^{(+)}$
is of rank 1, which signals the partial supersymmetry breaking:
\begin{eqnarray}
\llangle\delta
\left(\frac{\lambda^{\underline{i}}+\psi^{\underline{i}}}{\sqrt{2}}\right)
\rrangle
&=& im\sqrt{\frac{2}{N}} (\eta_1-\eta_2)~,\\
\llangle\delta
\left(\frac{\lambda^{\underline{i}}-\psi^{\underline{i}} }{\sqrt{2}}\right)
\rrangle
&=&
0~.
\end{eqnarray}
In the original Cartan basis, this means that 
$\llangle\delta\left(
\frac{\lambda^{i}+\psi^{i}}{\sqrt{2}}
\right)\rrangle=
2im\delta^i_0 (\eta_1-\eta_2)$
where we have used the fact
$\displaystyle{\sum_{\underline{j}}O_{\underline{j}}{}^i=\sqrt{2N}\delta_0^i}$.
As we will show in section VI,
the fermion $\frac{1}{\sqrt{2}}(\lambda^i+\psi^i)$
are massless while $\frac{1}{\sqrt{2}}(\lambda^i-\psi^i)$ are massive.
Thus, $\mathcal{N}=2$ supersymmetry is spontaneously broken to $\mathcal{N}=1$ 
and we obtain the Nambu-Goldstone fermion
$\frac{1}{\sqrt{2}}(\lambda^0+\psi^0)$ 
associated with the overall $U(1)$ part.
The same reasoning holds when $\boldsymbol{\mathcal{M}}_+$ is empty.
On the other hand, if neither $\boldsymbol{\mathcal{M}}_+$ 
nor $\boldsymbol{\mathcal{M}}_-$ is empty, 
we have two independent rank one matrices 
$\boldsymbol{\tau} \cdot \boldsymbol{d}^{(+)}$ and 
$\boldsymbol{\tau} \cdot \boldsymbol{d}^{(-)}$ and 
$\mathcal{N}=2$ supersymmetry is 
spontaneously broken to $\mathcal{N}=0$.
Which part of the Cartan subalgebra of $u(N)$ contains two Nambu-Goldstone fermions 
depend upon the type of grouping of $1 \sim N$
into $\boldsymbol{\mathcal{M}}_+$ and $\boldsymbol{\mathcal{M}}_-$.
Let $\underline{i}=(\underline{i'},\underline{i''})$
with 
$\underline{i'}\in\boldsymbol{\CM}_+$
and
$\underline{i''}\in\boldsymbol{\CM}_-$, then
\begin{eqnarray}
\llangle\delta\left(
\frac{\lambda^{\underline{i'}}+\psi^{\underline{i'}}}{\sqrt{2}}
\right)
\rrangle
&=&
im\sqrt{\frac{2}{N}}
(\eta_1-\eta_2)
~,~~~
\llangle\delta\left(
\frac{\lambda^{\underline{i''}}+\psi^{\underline{i''}}}{\sqrt{2}}
\right)
\rrangle
=0~,
\\
\llangle\delta\left(
\frac{\lambda^{\underline{i''}}-\psi^{\underline{i''}}}{\sqrt{2}}
\right)\rrangle
&=&
-im\sqrt{\frac{2}{N}}
(\eta_1+\eta_2)
~,~~~
\llangle\delta\left(
\frac{\lambda^{\underline{i'}}-\psi^{\underline{i'}}}{\sqrt{2}}
\right)\rrangle
=0
~.
\end{eqnarray}
As is obvious from the similar analysis given in section VI,
the fermions, 
$\frac{1}{\sqrt{2}}(\lambda^{\underline{i'}}+\psi^{\underline{i'}})$
and 
$\frac{1}{\sqrt{2}}(\lambda^{\underline{i''}}-\psi^{\underline{i''}})$,
are massless and contain two Nambu-Goldstone fermions
of $\CN=2$ supersymmetry broken to $\CN=0$. 

We comment on the vacuum value of the scalar potential $\CV$.
For the $\CN=1$ vacua, we have
\begin{eqnarray}
\llangle\CV\rrangle=\mp 2m\xi,
\end{eqnarray}
where the $\mp$ signs correspond respectively to the cases 
$\forall \underline{i}\in \boldsymbol{\CM}_\pm$.
For the $\CN=0$ vacua, on the other hand, we have
\begin{eqnarray}
\llangle\CV\rrangle=-2m\xi\frac{1}{N}
\left(
\textrm{ord}(\boldsymbol{\CM}_+)
- \textrm{ord}(\boldsymbol{\CM}_-)
\right)~,
\end{eqnarray}
where $\textrm{ord}(\boldsymbol{\CM}_{\pm})$ is the number of elements of 
$\boldsymbol{\CM}_{\pm}$.

We now impose the positivity criterion of 
the K\"ahler metric to select the physical vacua.
For the $\CN=1$ vacua, 
this criterion requires 
\begin{eqnarray}
\llangle g_{\underline{i}\underline{i}}\rrangle=\mp2\frac{\xi}{m} >0.
\end{eqnarray}
Depending upon $\frac{\xi}{m}\lessgtr 0$, 
we must choose either one of the two possibilities
discussed above.
For the $\CN=0$ vacua, however,
the K\"ahler metric cannot be positive definite,
causing
unconventional signs for the kinetic term.
The $\CN=0$ vacua are regarded as unphysical.

In the subsequent sections, we will analyse the $\mathcal{N}=1$ vacua.
We summarize some of the properties here in the original bases.
\begin{eqnarray}
\llangle \boldsymbol{D}_a \rrangle &=&
\delta_a^{\ i} \llangle \boldsymbol{D}_{i} \rrangle 
= \delta_a^{\ i} O_i^{\ \underline{j}} \llangle \boldsymbol{D}_{\underline{j}} \rrangle 
= \delta_a^{\ i} \sum_{\underline{j}} O_i^{\ \underline{j}} \boldsymbol{d}_{(\pm)}
=\delta_a^{\ i} \delta_i^{\ 0} \sqrt{\frac{N}{2}} \boldsymbol{d}_{\pm}
= \delta_a^{\ 0} \llangle \boldsymbol{D}_0 \rrangle \ .
~~~
\end{eqnarray}
Recalling $\partial_a W=e\delta_a^{\ 0}+m\mathcal{F}_{a0}$, 
we see
\begin{eqnarray}
\llangle \mathcal{F}_{a0} \rrangle=\delta_a^{\ 0} 
\llangle \mathcal{F}_{00} \rrangle \ ,
\ \ \ \ 
\llangle g_{a0} \rrangle=\delta_a^{\ 0} \llangle g_{00} \rrangle \ .
\end{eqnarray}
Hence
\begin{eqnarray}
0
&=&\llangle \boldsymbol{D}_0^* \cdot \boldsymbol{D}_0^* \rrangle
=2 \llangle (\partial_0 W)^2+\xi^2 \rrangle
=2(e+m\llangle \mathcal{F}_{00} \rrangle)^2+2 \xi^2, \\
\llangle \mathcal{F}_{00} \rrangle &=& 
- \left( \frac{e}{m} \pm i\frac{\xi}{m} \right), \label{F00} \\
\llangle \textrm{Re} \mathcal{F}_{00} \rrangle &=&
-\frac{e}{m}, \label{ReF00} \\
\llangle g_{00} \rrangle &=& 
\mp \frac{\xi}{m} = \left| \frac{\xi}{m} \right|. \label{g00}
\end{eqnarray}
After exhausting all possibilities, 
we conclude that partial spontaneous supersymmetry breaking 
takes place in the overall $U(1)$ sector.
As for $\llangle \mathcal{V} \rrangle$, we obtain 
\begin{eqnarray}
\llangle \mathcal{V} \rrangle=\mp 2m\xi=2|m\xi|.
\label{potential4.14}
\end{eqnarray}

\section{Gauge symmetry breaking} \label{GSB}

Following the analysis of the vacua of our model,
we turn to spontaneous breaking of gauge symmetry.

Let us recall that with the generic prepotential (\ref{prepot}),
the vacuum condition is given by (\ref{Fii:vacuum}):
\begin{eqnarray}
\llangle
\mathcal{F}_{\underline{i} \underline{i}}
\rrangle
+2 \zeta =0 \ , \label{4.1}
\end{eqnarray}
with
\begin{eqnarray}
\llangle
\mathcal{F}_{\underline{i} \underline{i}}
\rrangle
&=&\sum_{\ell}^{k=\deg \mathcal{F}} \frac{C_l}{(\ell-2)!} 
(\lambda^{\underline{i}})^{\ell-2} \nonumber \\
&=&C_2 +C_3 \lambda^{\underline{i}} +C_4 \frac{1}{2!} 
(\lambda^{\underline{i}})^2+C_5 \frac{1}{3!} (\lambda^{\underline{i}})^3+ 
\cdots \ \ . \label{Fii}
\end{eqnarray}
Here, we have introduced a complex parameter 
\begin{eqnarray}
\zeta \equiv \frac{e}{m} \pm i \frac{\xi}{m} \ .
\end{eqnarray}
Eq.(\ref{4.1}) is an algebraic equation for $\lambda^{\underline{i}}$ 
with degree $\deg \mathcal{F}-2=k-2$ and provides $k-2$ complex roots
denoted by $\lambda^{(\ell,\pm)}$, $\ell=1 \sim k-2$.
Thus each $\lambda^{\underline{i}}$ is determined to be one of these 
$k-2$ complex roots.
As is well-known, this defines a grouping of $N$ eigenvalues into 
$k-2$ sets and hence determines a breaking pattern of $U(N)$ gauge symmetry
into a product gauge group
$\displaystyle{\prod_{i=1}^{k-2} U(N_i)}$ with 
$\displaystyle{\sum_{i=1}^{k-2} N_i = N}$:
\begin{eqnarray}
\llangle A \rrangle =
\left(
\begin{array}{cccccccccc}
\lambda^{(1,\pm)} & & & & & & & & & \\
 & \ddots& & & & & & & & \\
 & & \lambda^{(1,\pm)}& & & & & & & \\
 & & & \lambda^{(2,\pm)}& & & & & & \\
 & & & & \ddots& & & & & \\
 & & & & & \lambda^{(2,\pm)}& & & & \\
 & & & & & & \ddots& & & \\
 & & & & & & & \lambda^{(k-2,\pm)}& & \\
 & & & & & & & & \ddots& \\
 & & & & & & & & & \lambda^{(k-2,\pm)}
\end{array}
\right) \ .
\end{eqnarray}
In fact, as we will see in the next section,
$v_m^\alpha$ are massless if 
$t_\alpha\in \{t_a ~|~ [t_a,\llangle A\rrangle]=0\}$
while $v_m^\mu$ are massive if 
$t_\mu \in \{t_a ~|~  [t_a,\llangle A\rrangle]\neq 0\}$.
The massless
 $v_m^\alpha$ contain gauge fields of unbroken 
$\displaystyle{\prod_iU(N_i)}$ and the superpartner
of the Nambu-Goldstone fermion lies in the overall $U(1)$ part.
We introduce 
\begin{eqnarray}
d_u \equiv \textrm{dim}\prod_i U(N_i).
\end{eqnarray}

Let us, as a warm up, work out the case $N=2$ and the case $N=3$.
\medskip

\noindent
1) $N=2$: there are two distinct cases for $\llangle A \rrangle$
\begin{eqnarray}
&& \textrm{i}) \ 
\left(
\begin{array}{cc}
\lambda^{(\pm)} & 0 \\
0 & \lambda^{(\pm)}
\end{array}
\right), \ \ \ \ 
\textrm{ii}) \ 
\left(
\begin{array}{cc}
\lambda^{(\pm)} & 0 \\
0 & {\lambda'}^{(\pm)}
\end{array}
\right) \ \textrm{with} \ \lambda^{(\pm)} \neq {\lambda'}^{(\pm)}. 
\label{caseofN=2}
\end{eqnarray}
The diagonal entries of $(\ref{caseofN=2})$ are chosen from 
$\lambda^{(\ell,\pm)}$, $\ell =1 \sim k-2$.
In respective cases, $\mathcal{N}=2$ supersymmetry and 
$U(2)$ gauge symmetry are broken respectively to
i) $\mathcal{N}=1$, $U(2)$ unbroken, ii) 
$\mathcal{N}=1$, $U(1) \times U(1)$.
\medskip

\noindent
2) $N=3$:
\begin{eqnarray}
&& \textrm{i}) \ 
\left(
\begin{array}{ccc}
\lambda^{(\pm)} & 0 & 0 \\
0 & \lambda^{(\pm)} & 0 \\
0 & 0 & \lambda^{(\pm)}
\end{array}
\right), \ \ \ \ 
\textrm{ii}) \ 
\left(
\begin{array}{ccc}
\lambda^{(\pm)} & 0 & 0 \\
0 & \lambda^{(\pm)} & 0 \\
0 & 0 & {\lambda'}^{(\pm)}
\end{array}
\right) \ \textrm{with} \ \lambda^{(\pm)} \neq {\lambda'}^{(\pm)}, 
\nonumber \\ 
&& \textrm{iii}) \ 
\left(
\begin{array}{ccc}
\lambda^{(\pm)} & 0 & 0 \\
0 & {\lambda'}^{(\pm)} & 0 \\
0 & 0 & {\lambda''}^{(\pm)} 
\end{array}
\right) \ \textrm{with} \ \lambda^{(\pm)} 
\neq {\lambda'}^{(\pm)} \neq {\lambda''}^{(\pm)} \neq \lambda^{(\pm)}, 
\nonumber 
\label{caseofN=3}
\end{eqnarray}
and the unbroken symmetries are respectively
i) $\mathcal{N}=1$, $U(3)$ unbroken, 
ii) $\mathcal{N}=1$, $U(2) \times U(1)$,
iii) $\mathcal{N}=1$, $U(1) \times U(1) \times U(1)$.

This pattern of symmetry breaking persists at general $N$.
Let $m_{\pm} \equiv \textrm{ord} (\boldsymbol{\mathcal{M}}_{\pm})$, namely,
the number of elements of $\boldsymbol{\mathcal{M}}_{\pm}$.
Because either $m_+=0$ or $m_-=0$, the unbroken symmetries are 
$\mathcal{N}=1$, as is 
established in the last section, and the product gauge groups 
$\displaystyle{\prod_{i=1}^n U(N_i)}$ with $n\leq k-2$, $N$.

Let us finally discuss the condition under which zero eigenvalues of 
$\left< A \right>$ emerge. This condition is simply 
\begin{eqnarray}
C_2 \pm 2 \zeta =0 \ , \label{zerocondition}
\end{eqnarray}
as is read off from eqs.(\ref{4.1}) and (\ref{Fii}).
The complex coefficients $C_{\ell}$ are to be determined from 
underlying microscopic theory and eq.(\ref{zerocondition})
tells that we can always finetune the single complex parameter $\zeta$ 
to obtain the zero eigenvalues. 
When $\mathcal{F}_{\underline{i} \underline{i}}$
is an even function of $\lambda^{\underline{i}}$, 
the condition means that eq.(\ref{4.1}) has a double root. 
As a prototypical example, let $\mathcal{F}_{\underline{i} \underline{i}}$
be even and $k=6$. The roots are 
\begin{eqnarray}
\lambda^{\underline{i}} = 
\pm \sqrt{\frac{-\frac{C_4}{2} \pm \sqrt{\left( \frac{C_4}{2} \right)^2
-4\cdot \frac{C_6}{4!}(C_2+2\zeta)}}{2\cdot \frac{C_6}{4!}}} ~.
\end{eqnarray}
When eq.(\ref{zerocondition}) is satisfied, 
one of the two pairs of complex roots coalesces and indeed develops 
into zero consisting of a double root.
This could be exploited to realize the \textquotedblleft triplet-doublet splitting" 
at $N=5$ in the context of $SU(5)$ $\mathcal{N}=1$ SGUT.
In the vacuum in which $U(5)$ is broken to $U(3)\times U(2)$ 
(or $SU(5)$ to $SU(3)\times SU(2)\times U(1)$, 
the standard model gauge group),
we are able to obtain
\begin{eqnarray}
\left< A \right>=
\left(
\begin{array}{ccccc}
\lambda& & & &0 \\
 &\lambda& & &  \\
 & &\lambda& &  \\
 & & &0&  \\
0& & & &0 \\
\end{array}
\right) \ ,
\end{eqnarray}
in the case in which degeneration of eigenvalues is favored.

\section{Mass spectrum} \label{sectionVI}

We examine the mass spectrum for the $\CN=1$
vacua for which $\forall \underline{i}\in \boldsymbol{\CM}_\pm$.

\subsection{Fermion mass spectrum} 
In this subsection,
we compute the fermion masses.
We examine the fermion mass term (\ref{L:mass})
 for $\psi^a$ and $\lambda^a$
\begin{eqnarray}
-\frac{i}{4}g^{cd^*}
\CF_{abc}
\partial_{d^*} W^*
 (\psi^a\psi^b+\lambda^a\lambda^b)
+\frac{1}{2\sqrt{2}}\left(
 g_{ac^*}k_b^*{}^{c}-g_{bc^*}k_a^*{}^{c}
 -\sqrt{2}\xi\delta_{c}^0
g^{cd}
\CF_{abd}
 \right) \psi^a\lambda^b
~.
 \label{fermion mass term}
\end{eqnarray}
The first term in (\ref{fermion mass term})
becomes
\begin{eqnarray}
-\frac{i}{4}\llangle g^{kl}\CF_{ijk}
\partial_{l^*}W^*
\rrangle
 (\psi^i\psi^j+\lambda^i\lambda^j)
-\frac{i}{4}\llangle g^{ij}\CF_{rsi}
\partial_{j^*}W^*
\rrangle
 (\psi^r\psi^s+\lambda^r\lambda^s)~,
 \label{fermion mass 1}
\end{eqnarray}
because
$\llangle\partial_{a^*}W^*\rrangle=
\delta_{a^*}^{i^*}\llangle\partial_{i^*}W^*\rrangle$
and $\llangle\CF_{ijr}\rrangle=0$.
The second term in (\ref{fermion mass term})
reduces to
\begin{eqnarray}
-\frac{1}{2}\llangle \xi g^{0k}\CF_{kij}\rrangle\psi^i\lambda^j
+\frac{1}{2\sqrt{2}}\llangle g_{st}f^t_{ri}A^{*i}\rrangle
(\psi^s\lambda^r-\psi^r\lambda^s)
-\frac{1}{2}\llangle \xi g^{0i}\CF_{irs}\rrangle\psi^r\lambda^s,
\label{fermion mass 2}
\end{eqnarray}
as
$ 
\llangle g_{ac^*}k^*_b{}^c\rrangle\psi^a\lambda^b
=\llangle g_{ac^*}f^c_{ri}A^{*i}\rrangle\psi^a\lambda^r
=\llangle g_{st}f^t_{ri}A^{*i}\rrangle\psi^s\lambda^r
$ 
. We have used $f^c_{ri} \llangle A^{*i} \rrangle=
f^t_{ri}\delta_t^c \llangle A^{*i} \rrangle$ 
in the last equality.
We thus conclude that the mass terms
of the fermions with $i$ index
are decoupled from those of the fermions with $r$ index.

{}From
(\ref{fermion mass 1}) and (\ref{fermion mass 2}),
 the mass term for the $\boldsymbol{\lambda}^{\underline{i}}_I$ 
in the eigenvalue basis 
is written as
$\frac{1}{2}\boldsymbol{\lambda}^{\underline{i}I}
(M_{\underline{i}\underline{i}})_I{}^{J}
\boldsymbol{\lambda}^{\underline{i}}_J$
with
\begin{eqnarray}
(M_{\underline{i}\underline{i}})_I{}^{J}=
-\frac{i}{2}\llangle 
g^{\underline{i}\underline{i}}\CF_{\underline{i}\underline{i}\underline{i}}
\rrangle
\left(
  \begin{array}{cc}
    -i\xi O^0_{\underline{i}}     &   
    \llangle \partial_{\underline{i}^*}W^* 
     \rrangle  \\
    -\llangle \partial_{\underline{i^*}}W^* 
     \rrangle   &
    +i\xi O^0_{\underline{i}}    \\
  \end{array}
\right)
=
\frac{1}{2\sqrt{2}}\llangle\CF_{\underline{i}\underline{i}\underline{i}}
\rrangle 
(\boldsymbol{\tau}\cdot \llangle \boldsymbol{D}^{\underline{i}} \rrangle)
~.
\end{eqnarray}
It is easy to 
show
that the mass term
becomes of the form
\begin{eqnarray}
&&
\frac{i}{8\sqrt{2}}
\sum_{\underline{i}}
\llangle D_2^{\underline{i}}+iD_3^{\underline{i}}\rrangle
\llangle \CF_{\underline{i}\underline{i}\underline{i}}
\rrangle
(\lambda^{\underline{i}}+\psi^{\underline{i}})^2
+\frac{i}{8\sqrt{2}}
\sum_{\underline{i}}
\llangle D_2^{\underline{i}}-iD_3^{\underline{i}}\rrangle
\llangle \CF_{\underline{i}\underline{i}\underline{i}}
\rrangle
(\lambda^{\underline{i}}-\psi^{\underline{i}})^2
~.
\end{eqnarray}
Noting that for the $\CN=1$ vacua, 
$\forall \underline{i}\in \boldsymbol{\CM}_\pm$,
\begin{eqnarray}
\llangle
D_2^{\underline{i}}\mp iD_3^{\underline{i}}
\rrangle
=
2i\frac{m}{\sqrt{N}}
~,~~~
\llangle
D_2^{\underline{i}}\pm iD_3^{\underline{i}}
\rrangle=0
~,
\end{eqnarray}
we find
that
 fermions
$\frac{1}{\sqrt{2}}(\lambda^{\underline{i}}\pm\psi^{\underline{i}})$
are massless 
while fermions
$\frac{1}{\sqrt{2}}(\lambda^{\underline{i}}\mp\psi^{\underline{i}})$
are massive with mass
$| m
\llangle g^{\underline{i} \underline{i}} \rrangle
\llangle 
\CF_{0\underline{i}\underline{i}}
\rrangle|$
after using (\ref{Ooj}).
Here $g^{\underline{i} \underline{i}}$ comes 
from the normalization of the kinetic term.

On the other hand,
it follows from (\ref{fermion mass 1})
and (\ref{fermion mass 2})
that the mass term of $\boldsymbol{\lambda}_I^r$ 
is written as
$\frac{1}{2}\boldsymbol{\lambda}^{rI}(M_{rs})_I{}^{J}\boldsymbol{\lambda}_J^s$
with
\begin{eqnarray}
&&
(M_{rs})_I{}^{J}=
\left(
  \begin{array}{cc}
 \mp m_{rs}+m'_{rs}    
&
 m_{rs}   \\  
 - m_{rs}   &
 \pm m_{rs}+m'_{rs}
\\
  \end{array}
\right)~,~~~~\\&&
m_{rs}=
- \frac{1}{2}
m
\llangle
 \mathcal{F}_{0 rs}
\rrangle~, \ \ \ 
m'_{rs}=\frac{1}{2\sqrt{2}}
\left(
\llangle g_{tr}\rrangle
f^t_{s\underline{i}}\lambda^{*\underline{i}}
-
\llangle g_{ts}\rrangle
f^t_{r\underline{i}}\lambda^{*\underline{i}}
\right)~.
\label{mass matrix:non Cartan}
\end{eqnarray}
It is convenient to express the index $r$
as a union of
 the two indices $r'$ and $\mu$
such that
$t_{r'}\in\{t_r~|~[t_r,\llangle A\rrangle]= 0\}$
and
$t_{\mu}\in\{t_r~|~[t_r,\llangle A\rrangle]\neq 0\}$.
Since $[t_{\mu},\llangle A\rrangle]$ belongs to $\{t_{\mu}\}$
and $\llangle g_{st}\rrangle$ is diagonal,
$m'_{rs}$ is nonvanishing only for $m'_{\mu\nu}$.
On the other hand, $m_{rs}$ is nonvanishing only for $m_{r's'}$, 
as $\llangle \mathcal{F}_{0 \mu \nu} \rrangle =0$.
This last equality is proven from (\ref{key relation}) and (\ref{4.1}) 
by
\begin{eqnarray}
\llangle 
\mathcal{F}_{0,\pm \underline{i}\underline{j},\pm \underline{i}\underline{j}}
\rrangle
=\frac{\llangle
\mathcal{F}_{\underline{i}\underline{i}}
\rrangle
-
\llangle
\mathcal{F}_{\underline{j}\underline{j}}
\rrangle}{2\sqrt{2N}(\lambda^{\underline{i}}-\lambda^{\underline{j}})}
=
\frac{-2\zeta
-(-2\zeta)
}{2\sqrt{2N}(\lambda^{\underline{i}}-\lambda^{\underline{j}})}=0~. 
\label{F0mumu=0}
\end{eqnarray}
Thus, we find that the $\boldsymbol{\lambda}_I^{r'}$ mass term 
decouples from the $\boldsymbol{\lambda}_I^{\mu}$ mass term.
The $\boldsymbol{\lambda}_I^{r'}$ mass term
can be easily diagonalized 
as was done for $\boldsymbol{\lambda}_I^{\underline{i}}$,
 and we find that
$\frac{1}{\sqrt{2}}(\lambda^{r'}\pm\psi^{r'})$
are massless and
$\frac{1}{\sqrt{2}}(\lambda^{r'}\mp\psi^{r'})$
are massive with mass
$|m \llangle g^{r'r'} \rrangle
\llangle
\CF_{0 r'r'}
\rrangle|$.
Combining the result on $\boldsymbol{\lambda}_I^{\underline{i}}$
with that on $\boldsymbol{\lambda}_I^{r'}$
and letting
$\alpha=\underline{i} \cup r'$ namely
$t_\alpha\in \{t_a~|~[t_a,\llangle A\rrangle]=0\}$,
we find that
$\frac{1}{\sqrt{2}}(\lambda^\alpha\pm\psi^\alpha)$
are massless while
$\frac{1}{\sqrt{2}}(\lambda^\alpha\mp\psi^\alpha)$
are massive with mass
$|m
\llangle
g^{\alpha\alpha}
\rrangle
\llangle
\CF_{0\alpha\alpha}
\rrangle|$. 

Finally, 
we examine the mass of
$\boldsymbol{\lambda}_I^{\mu}$.
Denoting $E^{\underline{i}\underline{j}}_\pm$ with 
$\lambda^{\underline{i}}\neq\lambda^{\underline{j}}$
by $t_{\mu_\pm}$ for short,
we find that 
$m'_{\mu \nu}$ is nonvanishing only for $m'_{\mu_+ \mu_-}=-m'_{\mu_- \mu_+}$,
since $[\llangle A\rrangle,t_{\mu_\pm}]\propto t_{\mu_\mp}$.
(See the Appendix.)
In these indices,
the $\boldsymbol{\lambda}_I^{\mu_{\pm}}$ mass term 
is written as
\begin{eqnarray}
\frac{1}{2} \boldsymbol{\lambda}^{\mu_+ I}
\left(
\begin{array}{cc}
2m'_{\mu_+ \mu_-} & 0 \\
 0                & 2m'_{\mu_+ \mu_-}
\end{array}
\right)_I^{\ J}
\boldsymbol{\lambda}_J^{\mu_-}. \label{massoflambda}
\end{eqnarray}
The summation is implied only for either one of the two indices 
$\mu_+$ or $\mu_-$, 
and it is over half of the broken generators.
This implies that 
$\boldsymbol{\lambda}_I^{\mu}$
has mass 
$
\left|
g^{\nu\nu}m_{\mu\nu}'
\right|=
\left|
g^{\mu_-\mu_{-}}m'_{\mu_+ \mu_-}
\right|
=
\frac{1}{\sqrt{2}} 
\left|
f_{\mu \underline{i}}^{\nu} \lambda^{*\underline{i}}
\right|
$
where we have used the fact 
$\llangle g_{\mu_+ \mu_+} \rrangle= 
\llangle g_{\mu_- \mu_-} \rrangle$
and 
$f_{\mu_+ \underline{i}}^{\mu_-} \lambda^{*\underline{i}}=
-f_{\mu_- \underline{i}}^{\mu_+} \lambda^{*\underline{i}}$.

\subsection{Boson mass spectrum}

In order to obtain the boson masses of our model, 
we evaluate the second variation of the scalar potential
on the vacuum $\llangle \delta \delta \mathcal{V} \rrangle$.
Recall that 
\begin{eqnarray}
\mathcal{V}&=&
\mathcal{V}_1+\mathcal{V}_2, \\
\mathcal{V}_1&=&
\frac{1}{8} g_{ab} \mathfrak{D}^a \mathfrak{D}^b ,
\ \ \mathfrak{D}^a=-i f_{cd}^a A^{*c} A^d, \\
\mathcal{V}_2&=&
g^{ab} \left( \xi^2 \delta_a^{\ 0} \delta_b^{\ 0}+
\partial_a W \partial_b W^* \right) 
\nonumber \\ &=& 
(\xi^2 +e^2) g^{00}+m^2 g_{00}+2me(\textrm{Re} \mathcal{F}_{0b}) g^{b0} 
+m^2 (\textrm{Re} \mathcal{F}_{0b}) g^{bc} \textrm{Re} \mathcal{F}_{0c} \ .
\end{eqnarray}
Let us first compute $\llangle \delta \delta \mathcal{V}_1 \rrangle$. From 
$\llangle \mathfrak{D}^a \rrangle=0$ and 
$\llangle A^a \rrangle=\delta^a_{\ j} \llangle A^j \rrangle$, we obtain
\begin{eqnarray}
\llangle \delta \mathfrak{D}^a \rrangle&=&
\delta^a_{\ \tilde{\mu}} 
\llangle \delta \mathfrak{D}^{\tilde{\mu}} \rrangle \ , \\
\llangle \delta \mathfrak{D}^{\tilde{\mu}} \rrangle &=& 
\llangle \delta \mathfrak{D}^{\tilde{\mu}} \rrangle ^*
=
\left(
(-i)f^{\tilde{\mu}}_{\underline{j}\mu} 
\lambda^{*\underline{j}},\ (+i)f^{\tilde{\mu}}_{\underline{j} \mu} 
\lambda^{\underline{j}}
\right)
\left(
\begin{array}{c}
\delta A^{\mu} \\
\delta A^{*\mu}
\end{array}
\right)
\equiv
\left(
\overrightarrow{
(f_{\bot } \lambda^*)
^{\tilde{\mu}}_{\ \mu}
}
\right)^t
\cdot
\overrightarrow{\delta A^{\mu}} \ .
~~~
\end{eqnarray}
Here $f^{\tilde{\mu}}_{\underline{j}\mu}$ is the structure constant of 
the $u(N)$ Lie algebra
which is read off from the Appendix. 
For given $\mu$ specified by a pair of indices 
$(\underline{k},\underline{l}),\ 1\leq \underline{k} \leq \underline{l} \leq N$,
$\tilde{\mu}$ is uniquely determined and vice versa and 
the summation over $\underline{j}$ reduces to that of 
$\underline{j}=\underline{k}$ and $\underline{j}=\underline{l}$.
We obtain
\begin{eqnarray}
\llangle \delta \delta \mathcal{V}_1 \rrangle 
=\frac{1}{4} \sum_{\tilde{\mu}} 
\llangle \delta \mathfrak{D}^{\tilde{\mu}} \rrangle
\llangle g_{\tilde{\mu} \tilde{\mu}} \rrangle
\llangle \delta \mathfrak{D}^{\tilde{\mu}} \rrangle. \label{deldelV1}
\end{eqnarray}
The summation over $\tilde{\mu}$ is for $N^2-d_u$ directions of 
the broken generators.

In order to separate the $N^2-d_u$ Nambu-Goldstone zero modes 
(one mode for each $\mu$),
we introduce the following projector of a $2\times 2$ matrix 
as a diad for each $\mu$:
\begin{eqnarray}
\mathcal{P}^{\tilde{\mu}}_{\ \mu}
\equiv
\frac{1}
{||
\overrightarrow{(f_{\bot} \lambda)^{\tilde{\mu}}_{\ \mu}
}||^2} 
\overrightarrow{(f_{\bot}\lambda)^{\tilde{\mu}}_{\ \mu}} 
\left(\overrightarrow{(f_{\bot}\lambda^*)^{\tilde{\mu}}_{\ \mu}}\right)^t,
\end{eqnarray}
where
\begin{eqnarray}
||\overrightarrow{(f_{\bot}\lambda)^{\tilde{\mu}}_{\ \mu}}||^2
&\equiv&
\left( \overrightarrow{(f_{\bot}\lambda^*)^{\tilde{\mu}}_{\ \mu}} \right)^t
\cdot
\overrightarrow{(f_{\bot}\lambda)^{\tilde{\mu}}_{\ \mu}} \nonumber \\
&=& 2|f^{\tilde{\mu}}_{\underline{j}\mu} \lambda^{\underline{j}}|^2.
\end{eqnarray}
We express $\llangle \delta \mathfrak{D}^{\tilde{\mu}} \rrangle$ as
\begin{eqnarray}
\llangle \delta \mathfrak{D}^{\tilde{\mu}} \rrangle
=
\left(
\overrightarrow{
(f_{\bot}\lambda^*)^{\tilde{\mu}}_{\ \mu}
}
\right)^t
\cdot
\left(
\mathcal{P}^{\tilde{\mu}}_{\ \mu}
\overrightarrow{\delta A^{\mu}}
\right) \ . \label{delmathfrakD}
\end{eqnarray}
The mode orthogonal to 
$\left(\mathcal{P}^{\tilde{\mu}}_{\ \mu} \overrightarrow{\delta A^{\mu}} \right)$
is the one belonging to the zero eigenvalue for each $\mu$ in eq.(\ref{deldelV1}) 
and is absorbed into the longitudinal components of the corresponding gauge fields.
It is given by
\begin{eqnarray}
(\boldsymbol{1}_2-\mathcal{P}^{\tilde{\mu}}_{\ \mu})
\overrightarrow{\delta A^{\mu}},
\end{eqnarray}
with
\begin{eqnarray}
\boldsymbol{1}_2-\mathcal{P}^{\tilde{\mu}}_{\ \mu}
=
\frac{1}{|| \overrightarrow{(f_{\bot}\lambda)^{\tilde{\mu}}_{\mu}} ||^2}
\left(
\begin{array}{c}
f^{\tilde{\mu}}_{\underline{j}\mu} \lambda^{\underline{j}} \\
f^{\tilde{\mu}}_{\underline{j}\mu} \lambda^{*\underline{j}}
\end{array}
\right)
\left(
f^{\tilde{\mu}}_{\underline{j}\mu} \lambda^{*\underline{j}},
\ 
f^{\tilde{\mu}}_{\underline{j}\mu} \lambda^{\underline{j}} 
\right) \ .
\end{eqnarray}
Substituting (\ref{delmathfrakD}) into (\ref{deldelV1}), we obtain
\begin{eqnarray}
\llangle \delta \delta \mathcal{V}_1 \rrangle 
=
\frac{1}{4} \sum_{\tilde{\mu}} \llangle g_{\tilde{\mu} \tilde{\mu}} \rrangle
|f^{\tilde{\mu}}_{\underline{j} \mu} \lambda^{\underline{j}}|^2
\left(
\mathcal{P}^{\tilde{\mu}}_{\ \mu} \overrightarrow{\delta A^{\mu}}
\right)^{*t}
\cdot
\left(
\mathcal{P}^{\tilde{\mu}}_{\ \mu} \overrightarrow{\delta A^{\mu}}
\right) \ .
\label{deldelV1-2}
\end{eqnarray}

Let us turn our attention to $\llangle \delta \delta \mathcal{V}_2 \rrangle$.
\begin{eqnarray}
\delta \mathcal{V}_2 &=&
-(\xi^2+e^2)(g^{-1}\delta g g^{-1})^{00} 
+ m^2 (\delta g)_{00} \nonumber \\
&&+2me (\delta \textrm{Re} \mathcal{F}_{0b})g^{b0}
-2me (\textrm{Re} \mathcal{F}_{0b})(g^{-1}\delta g g^{-1})^{b0} \nonumber \\
&&+2m^2 (\delta \textrm{Re} \mathcal{F}_{0b})g^{bc} \textrm{Re} \mathcal{F}_{0c}
-m^2 (\textrm{Re} \mathcal{F}_{0b})(g^{-1}\delta g g^{-1})^{bc} 
(\textrm{Re} \mathcal{F}_{0c}) \ .
\end{eqnarray}
It is easy to check $\llangle \delta \mathcal{V}_2 \rrangle=0$, 
as it should be, from the properties 
(\ref{F00})-(\ref{g00}) listed in the end of section 
\ref{PBSsection}
It is straightforward to carry out one more variation, 
which we will not spell out here.
Again from (\ref{F00})-(\ref{g00}), we obtain
\begin{eqnarray}
\llangle
\delta \delta \mathcal{V}_2
\rrangle
&=&2m^2 \llangle \delta \mathcal{F}^*_{0c} \rrangle \llangle g^{cd} \rrangle 
\llangle \delta \mathcal{F}_{0d} \rrangle \nonumber \\
&=&2m^2 \delta A^{*b} \llangle \mathcal{F}_{0bc}^* \rrangle \llangle g^{cd} \rrangle 
\llangle \mathcal{F}_{0ad} \rrangle \delta A^a \nonumber \\
&=&2m^2 \delta A^{*\alpha} \llangle \mathcal{F}_{0\alpha \alpha}^* \rrangle 
\llangle g^{\alpha \alpha} \rrangle \llangle \mathcal{F}_{0\alpha \alpha} \rrangle 
\delta A^{\alpha}
\ .
\label{deldelV2}
\end{eqnarray}
In deriving the last line, we have used eq.(\ref{F0mumu=0}) 
derived in the preceding subsection.

Combining the two calculations (\ref{deldelV1-2}) and (\ref{deldelV2}),
we can read off the mass formula of the scalar bosons from 
$\llangle \delta \delta \mathcal{V} \rrangle=
\llangle \delta \delta \mathcal{V}_1 \rrangle+
\llangle \delta \delta \mathcal{V}_2 \rrangle$.
The scalar masses arising from the directions of the unbroken 
generators are given by 
$
|m \llangle g^{\alpha \alpha} \rrangle 
\llangle \mathcal{F}_{0\alpha \alpha}\rrangle |
$.
Here, we have interpreted $\llangle g^{\alpha \alpha} \rrangle$ 
as a factor due to the normalization of the kinetic term.
The scalar masses arising from the directions of the broken generators 
are given by 
$\frac{1}{\sqrt{2}}|f^{\tilde{\mu}}_{\underline{j}\mu}\lambda^{\underline{j}}|$,
after the normalization we just discussed.
Finally we read off the mass of the massive gauge bosons from 
$-\llangle \mathcal{L}_{\textrm{kin}} \rrangle$.
\begin{eqnarray}
-\llangle \mathcal{L}_{\textrm{kin}} \rrangle 
&=& \llangle g_{aa'} \rrangle \frac{1}{2} f_{bc}^a v_m^{\ b} 
\llangle A^c \rrangle \frac{1}{2} f_{b'c'}^{a'} v^{mb'} \llangle A^{c'} \rrangle 
\nonumber \\
&=& \frac{1}{4} \sum_{\mu} 
|f^{\tilde{\mu}}_{\underline{j}\mu}\lambda^{\underline{j}}|^2 v_m^{\ \mu} v^{m\mu}.
\end{eqnarray}
The mass is given by 
$\frac{1}{\sqrt{2}}
|f^{\tilde{\mu}}_{\underline{j} \mu}\lambda^{\underline{j}}|
$.

We summarize the spectrum of the bosons and the fermions
by the following table:
\begin{eqnarray}
  \begin{array}{c|c|c|c|c}
  
      & \textrm{field}   & \textrm{mass}  &{\rm label} & 
      \textrm{\# of polarization states} \\
    \hline
  
       &v_m^\alpha    & 0   &A & 2d_u  \\
    \hline
  
       & v_m^\mu   &\frac{1}{\sqrt{2}}
|f^\nu_{\mu \underline{i}}\lambda^{\underline{i}}|    &C &3(N^2-d_u)\\
  
    \hline
       &\frac{1}{\sqrt{2}}(\lambda^\alpha\pm\psi^\alpha)    &0    &A &2d_u\\
  
    \hline
       &\frac{1}{\sqrt{2}}(\lambda^{\alpha}\mp\psi^{\alpha})   
        & |{m} \llangle g^{\alpha \alpha}\rrangle 
        \llangle 
\CF_{0\alpha\alpha}
\rrangle|    &B &2d_u\\  
    \hline
       &\boldsymbol{\lambda}^{\mu}_I    
       & \frac{1}{\sqrt{2}}|
       f^{\nu}_{\mu \underline{i}}\lambda^{\underline{i}}|  & C &4(N^2-d_u)\\     
    \hline
       &A^{{\alpha}}    
       &
|m \llangle g^{\alpha \alpha}\rrangle
\llangle
\CF_{0\alpha\alpha}
\rrangle|    &B &2d_u\\
    \hline
       &\mathcal{P}_{\mu}^{\tilde{\mu}} A^\mu    &\frac{1}{\sqrt{2}}| 
       f^{\nu}_{\mu \underline{i}}\lambda^{\underline{i}}|  &C &N^2-d_u\\
  \end{array}
\nonumber
\end{eqnarray}

We find the following supermultiplets.
First,
we consider the massless particles associated with 
$\frac{1}{\sqrt{2}}(\lambda^\alpha\pm\psi^\alpha)$
and
$v_m^\alpha$.
These are labelled as A in the table.
These
form massless $\CN=1$ vector multiplets
of spin $(1/2,1)$,
in which the Nambu-Goldstone vector multiplet is contained
in the overall $U(1)$ part.
Second, massive particles labelled as B, 
which are associated with 
$\frac{1}{\sqrt{2}}(\lambda^{\alpha}\mp\psi^{\alpha})$
and $A^{\alpha}$,
have masses given by  
$|m\llangle
g^{\alpha\alpha}
\rrangle
\llangle \CF_{0\alpha \alpha} \rrangle|$.
These form massive $\CN=1$ chiral multiplets of spin $(0,1/2)$.
Finally, we consider massive particles labelled as C.
The zero modes of $A^{\mu}$ are absorbed into $v_m^{\mu}$
as the longitudinal modes to form massive vector fields.
These form two massive multiplets of spin $(0,1/2,1)$ with 
the massive modes of $A^{\mu}$ and $\boldsymbol{\lambda}^{\mu}_I$.
The masses of these supermultiplets are given by
$\frac{1}{\sqrt{2}}\left|
f^{\nu}_{\mu \underline{i}}\lambda^{\underline{i}}\right|$.

In the following figure, 
the masses of the three types of $\mathcal{N}=1$ 
supermultiplets
are 
schematically drawn.

\begin{figure}[h]
   \begin{center}
      \psfig{file=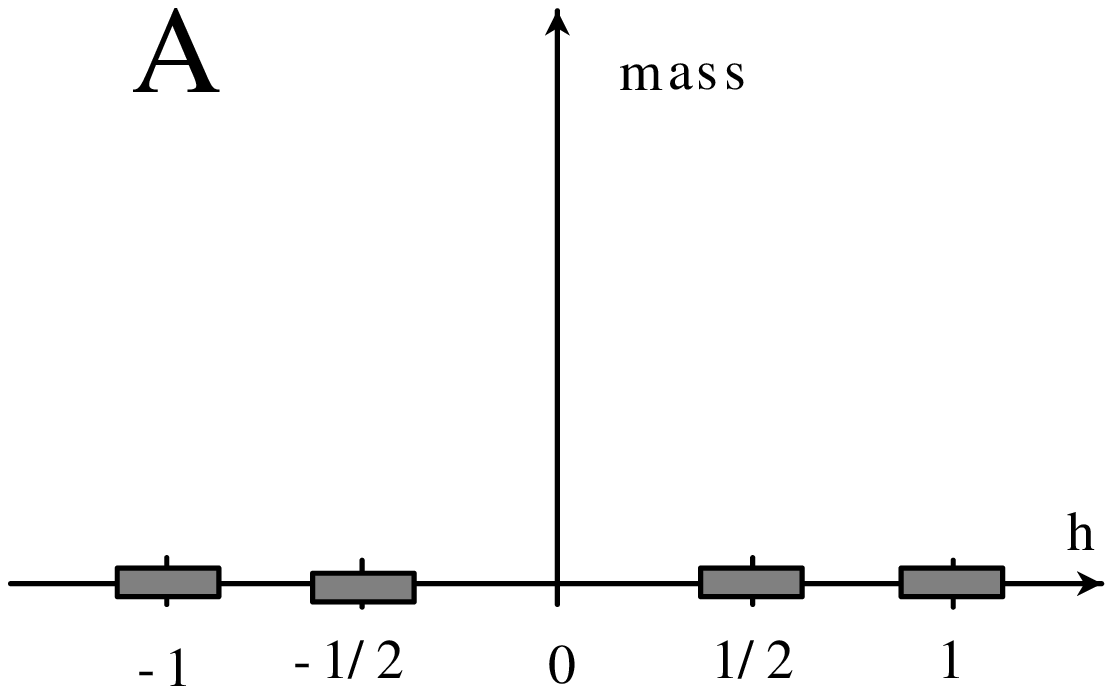,height=30mm}
      \psfig{file=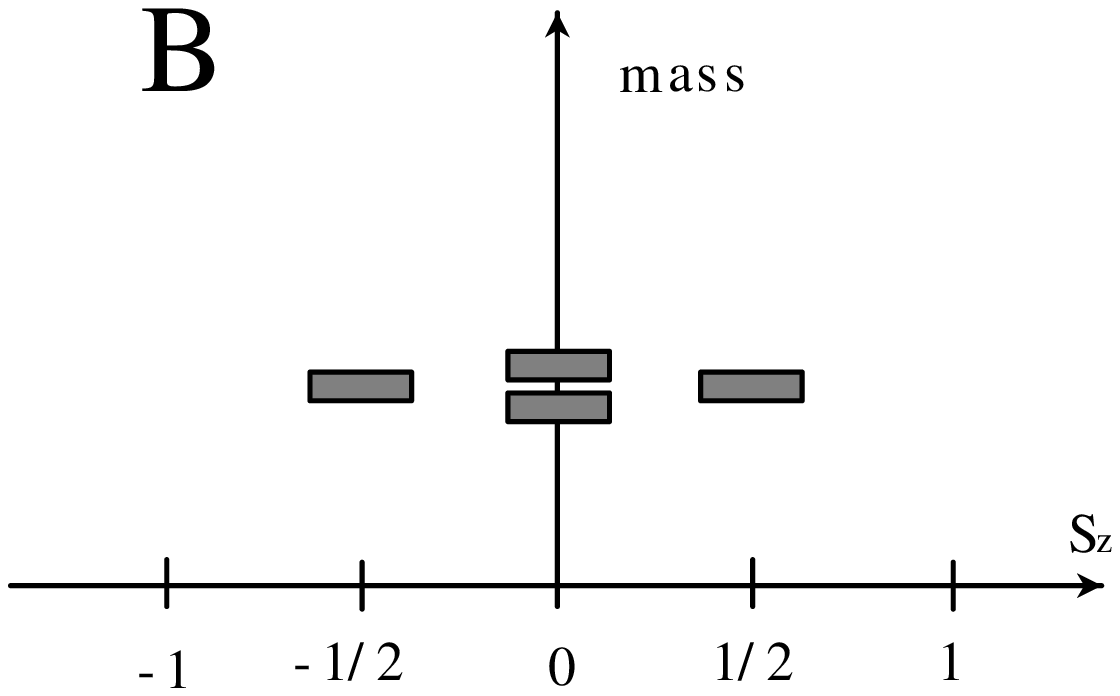,height=30mm}
      \psfig{file=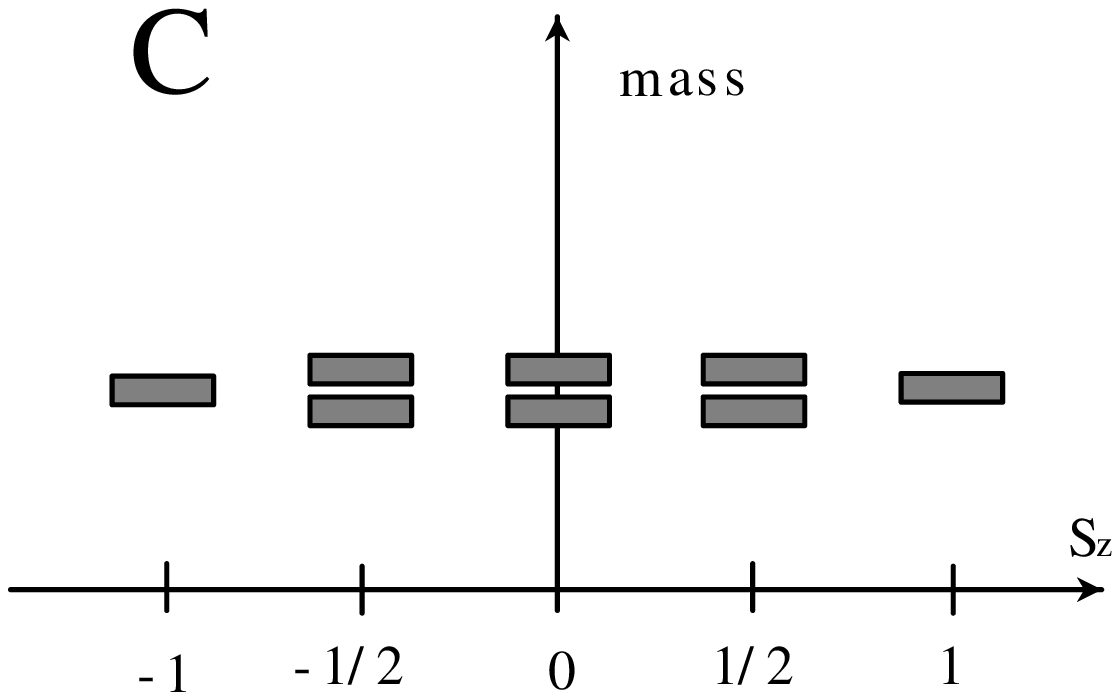,height=30mm} 
   \end{center}
\end{figure}

\newpage

\section*{Acknowledgements}
The authors thank 
Koichi Murakami
and Yukinori Yasui
for useful discussions.
This work is supported in part by the Grant-in-Aid for Scientific
Research(16540262) from the Ministry of Education,
Science and Culture, Japan.
Support from the 21 century COE program
``Constitution of wide-angle mathematical basis focused on knots"
is gratefully appreciated.
The preliminary version of this work was presented 
at the international workshop
``Frontier of Quantum Physics''
in the Yukawa Institute
for Theoretical Physics, Kyoto University (February 17-19 2005).
We wish to acknowledge the participants
for stimulating discussions.

\appendix
\section{}
Let $E_{\underline{i}\underline{j}}$, 
$\underline{i},\underline{j}=\underline{1},...,\underline{N}$, 
be the fundamental matrix,
which has $1$ at the $(\underline{i},\underline{j})$-component
and $0$ otherwise.
Cartan generators of $u(N)$ are 
$H_{\underline{i}}\equiv E_{\underline{i}\underline{i}}$,
which are denoted as $t_{\underline{i}}$ in section 3.
Non-Cartan generators can be written as
\begin{eqnarray}
E_{\underline{i}\underline{j}}^+=
E_{\underline{j}\underline{i}}^+
\equiv \frac{1}{2}(E_{{\underline{i}}{\underline{j}}}+
E_{{\underline{j}}{\underline{i}}})
~,~~~
E_{{\underline{i}}{\underline{j}}}^-=
-E_{{\underline{j}}{\underline{i}}}^-
\equiv -\frac{i}{2}( E_{{\underline{i}}{\underline{j}}}-
E_{{\underline{j}}{\underline{i}}})
~,~~~~~~{\underline{i}}\neq {\underline{j}}~,
\end{eqnarray}
which are normalized as 
$\tr(E_{{\underline{i}}{\underline{j}}}^\pm)^2=\frac{1}{2}$.
Commutation relations are
\begin{eqnarray}
{[}H_{\underline{i}},E_{{\underline{j}}{\underline{k}}}^\pm{]}&=&
\pm i\delta_{{\underline{i}}{\underline{j}}}E^\mp_{{\underline{i}}{\underline{k}}}+ 
i\delta_{{\underline{i}}{\underline{k}}}E^\mp_{{\underline{i}}{\underline{j}}}~,\\
{[}E^\pm_{{\underline{i}}{\underline{j}}},E^\pm_{{\underline{k}}\underline{l}}{]}&=&
\pm 2i\delta_{{\underline{j}}{\underline{k}}}E_{{\underline{i}}\underline{l}}^-
+2i\delta_{{\underline{i}}{\underline{k}}}E_{{\underline{j}}\underline{l}}^- 
+ 2i\delta_{{\underline{j}}\underline{l}}E_{{\underline{i}}{\underline{k}}}^- 
\pm 2i\delta_{{\underline{i}}\underline{l}}E_{{\underline{j}}{\underline{k}}}^-
~,\\
{[}E^+_{{\underline{i}}{\underline{j}}},E^-_{{\underline{k}}\underline{l}}{]}&=&
- 2i\delta_{{\underline{j}}{\underline{k}}}(E_{{\underline{i}}\underline{l}}^+
+\delta_{{\underline{i}}\underline{l}}H_{\underline{i}}) 
- 2i\delta_{{\underline{i}}{\underline{k}}}(E_{{\underline{j}}\underline{l}}^+
+\delta_{{\underline{j}}\underline{l}}H_{\underline{j}})
\nonumber\\&& 
+ 2i\delta_{{\underline{j}}{\underline{l}}}(E_{{\underline{i}}{\underline{k}}}^+
+\delta_{{\underline{i}}{\underline{k}}}H_{\underline{i}}) 
+ 2i\delta_{{\underline{i}}{\underline{l}}}(E_{{\underline{j}}{\underline{k}}}^+
+\delta_{{\underline{j}}{\underline{k}}}H_{\underline{j}}) 
~.
\end{eqnarray}
By introducing the vacuum expectation value 
$\langle\Phi\rangle=\lambda^{\underline{i}}t_{\underline{i}}$,
it follows 
that
\begin{eqnarray}
{[}E_{\underline{j}\underline{k}}^\pm,\langle A\rangle {]}&=&
{[}E_{{\underline{j}}{\underline{k}}}^\pm,\lambda^{\underline{i}} 
H_{\underline{i}}{]}=
\mp i(\lambda^{\underline{j}}-\lambda^{\underline{k}})
E_{{\underline{j}}{\underline{k}}}^\mp~.
\label{[E,A]}
\end{eqnarray}
In the text, $E_{{\underline{j}}{\underline{k}}}^\pm$ are denoted as $t_{r'}$ 
when $\lambda^{\underline{j}}=\lambda^{\underline{k}}$
while $t_{\mu}$ when $\lambda^{\underline{j}}\neq\lambda^{\underline{k}}$.
Explicitly, $\Phi=\Phi^at_a$ is expanded as follows:
\begin{eqnarray}
\Phi&=&\Phi^it_i+\Phi^rt_r~,\nonumber\\
\Phi^it_i&=&\Phi^{\underline{i}}t_{\underline{i}}~,~~~
\Phi^rt_r=\Phi^{r'}t_{r'}+\Phi^{\mu}t_{\mu}~,\nonumber\\
\Phi^{r'}t_{r'}&=&\left[
\frac{1}{2}\Phi^{\underline{i}\underline{j}}_+E_{\underline{i}\underline{j}}^+
+\frac{1}{2}\Phi^{\underline{i}\underline{j}}_-E_{\underline{i}\underline{j}}^-
\right]_{\lambda^{\underline{i}}=\lambda^{\underline{j}}}~,~~~
\Phi^{\mu}t_{\mu}=\left[
\frac{1}{2}\Phi^{\underline{i}\underline{j}}_+E_{\underline{i}\underline{j}}^+
+\frac{1}{2}\Phi^{\underline{i}\underline{j}}_-E_{\underline{i}\underline{j}}^-
\right]_{\lambda^{\underline{i}}\neq\lambda^{\underline{j}}}~.
\end{eqnarray}
Let $\CF$ be $\displaystyle{\CF=\sum_\ell \frac{C_\ell}{\ell!}\tr\Phi^\ell}$
then
$\langle\CF_{ab}\rangle$ are evaluated as
\begin{eqnarray}
\langle\CF_{{\underline{i}}{\underline{j}}}\rangle&=&
\sum_\ell \frac{C_\ell}{(\ell-1)!}
\sum_{l'=0}^{\ell-2}\tr(H_{{\underline{i}}}\langle A\rangle^l H_{{\underline{j}}}
\langle A\rangle^{\ell-2-l'})
\nonumber\\&=&
\sum_\ell \frac{C_\ell}{(\ell-1)!}
\sum_{l'=0}^{\ell-2}\tr(H_{{\underline{i}}}H_{{\underline{i}}'} H_{{\underline{j}}}
H_{{\underline{j}}'})
(\lambda^{{\underline{i}}'})^{l'}(\lambda^{{\underline{j}}'})^{\ell-2-l'} 
\nonumber \\
&=&\sum_\ell \frac{C_\ell}{(\ell-2)!}
\delta_{{\underline{i}}{\underline{j}}}
(\lambda^{\underline{i}})^{\ell-2}~,\\
\langle\CF_{\pm {\underline{i}}{\underline{j}}, 
\pm {\underline{i}}{\underline{j}}}\rangle&=&
\langle\frac{\partial^2\CF}
{\partial{\Phi^{{\underline{i}}{\underline{j}}}_\pm}
\partial{\Phi^{{\underline{i}}{\underline{j}}}_\pm}}
\rangle
=
\sum_\ell \frac{C_\ell}{(\ell-1)!}
\sum_{l'=0}^{\ell-2}\tr(E_{{\underline{i}}{\underline{j}}}^\pm 
H_{{\underline{i}}'}
E_{{\underline{i}}{\underline{j}}}^\pm H_{{\underline{j}}'}
)
(\lambda^{{\underline{i}}'})^{l'}
(\lambda^{{\underline{j}}'})^{\ell-2-l'} \nonumber\\
&=&
\left\{
\begin{array}{l}\displaystyle 
\sum_\ell \frac{C_\ell}{2(\ell-1)!} 
\frac{(\lambda^{\underline{i}})^{\ell-1}-(\lambda^{\underline{j}})^{\ell-1}}
{\lambda^{\underline{i}}-\lambda^{\underline{j}}} \ \ 
(\lambda^{\underline{i}} \neq \lambda^{\underline{j}} \textrm{~:~broken}) \\
\displaystyle 
\sum_\ell \frac{C_\ell}{2(\ell-2)!} (\lambda^{\underline{i}})^{\ell-2} 
\ \ \ \ \ \ \ \ \ \ 
(\lambda^{\underline{i}} = \lambda^{\underline{j}} \textrm{~:~unbroken})
\end{array}
\right.
\end{eqnarray}
and the others vanish.
We have used (\ref{[E,A]}) and
\begin{eqnarray}
E^\pm_{{\underline{i}}{\underline{j}}}E^\pm_{{\underline{k}}{\underline{l}}}&=&
\frac{1}{4}(\delta_{{\underline{i}}{\underline{k}}}
\delta_{{\underline{j}}{\underline{l}}}
\pm \delta_{{\underline{i}}{\underline{l}}}
\delta_{{\underline{j}}{\underline{k}}})
(H_{\underline{i}}+H_{\underline{j}}) 
+ \mbox{non-Cartan generators}~,
\\
E^\pm_{{\underline{i}}{\underline{j}}}E^\mp_{{\underline{k}}{\underline{l}}}&=&
\pm\frac{i}{4}(\delta_{{\underline{i}}{\underline{k}}}
\delta_{{\underline{j}}{\underline{l}}}
\mp \delta_{{\underline{i}}{\underline{l}}}\delta_{{\underline{j}}{\underline{k}}})
(H_{\underline{i}}-H_{\underline{j}}) 
+ \mbox{non-Cartan generators}~ 
\end{eqnarray}
in the calculation.

\newpage

\end{document}